\documentclass[12pt,preprint]{aastex}
\usepackage{tipa}
\usepackage{txfonts}
\usepackage{amssymb}

\begin{document}

\title{ON THE STRUCTURE OF ACCRETION DISKS WITH OUTFLOWS}

\author {Cheng-Liang Jiao\thanks{E-mail: chengliang.jiao@pku.edu.cn}  \ and   Xue-Bing Wu\thanks{E-mail: wuxb@pku.edu.cn}\\
 Department of Astronomy, School of Physics, Peking
University, Beijing 100871, China}

\begin{abstract}
In order to study the outflows from accretion disks, we solve the
set of hydrodynamic equations for accretion disks in the spherical
coordinates ($r\theta\phi$) to obtain the explicit structure along
the $\theta$ direction. Using self-similar assumptions in the radial
direction, we change the equations to a set of ordinary differential
equations (ODEs) about the $\theta$-coordinate, which are then
solved with symmetrical boundary conditions in the equatorial plane,
and the velocity field is obtained. The $\alpha$ viscosity
prescription is applied and an advective factor $f$ is used to
simplify the energy equation.The results display thinner,
quasi-Keplerian disks for Shakura-Sunyaev Disks (SSDs) and thicker,
sub-Keplerian disks for Advection Dominated Accretion Flows (ADAFs)
and slim disks, which are consistent with previous popular
analytical models. However, an inflow region and an outflow region
always exist, except when the viscosity parameter $\alpha$ is too
large, which supports the results of some recent numerical
simulation works. Our results indicate that the outflows should be
common in various accretion disks and may be stronger in slim disks,
where both advection and radiation pressure are dominant. We also
present the structure dependence on the input parameters and discuss
their physical meanings. The caveats of this work and possible
improvements in the future are discussed.

\end{abstract}


\section{INTRODUCTION}

The accretion disk models have been developed much in the past
several decades. Many disk models have been proposed and some of
them are widely adopted in astrophysical studies nowadays, including
but not restricted to Shakura-Sunyaev Disk (SSD; see Shakura \&
Sunyaev 1973), Advection Dominated Accretion Flow (ADAF; see Narayan
\& Yi 1994; Abramowicz et al. 1995), and slim disk (Abramowicz et
al. 1988). However, the outflow structure of accretion disks still
remains an open problem. The equations that describe the
hydrodynamic processes are the Navier-Stokes equations, which are
quite difficult to solve in the case of accretion disks which
involve viscosity and radiation. Therefore, in most works, some kind
of simplification, such as one-zone or polytropic distribution and
hydrostatic equilibrium, are usually applied in the vertical
direction, and the vertical variation ($z$ dependence in cylindrical
coordinates) of the velocity field is usually neglected. In this way
the equations are changed to a set of ordinary differential
equations (ODEs) in the radial direction, which can be solved
numerically. However, by taking these assumptions, one cannot get a
clear picture of the vertical structure of accretion flows; the
velocity is always radially inward and no mass will cross the disk
surface, displaying no outflow structure. Among the exceptions is a
work done by Narayan \& Yi in 1995(hereafter NY95), which used
self-similar assumptions in the radial direction and solved the
structure along the $\theta$ direction in spherical  coordinates
($r\theta\phi$). However, in their work they assumed $v_\theta=0$
and thus cannot get a proper velocity field, and their solutions
compose of only pure inflow. They argued that the Bernoulli
parameter is positive in their solutions so that a bipolar outflow
is expected to develop near the vertical axis. Blandford \& Begelman
(1999, hereafter BB99) relaxed the mass conservation assumption and
assumed that the mass inflow rate varies with radius, and obtained
solutions with outflow called adiabatic inflow-outflow solutions
(ADIOS). Their solutions are one-dimensional self-similar solutions
that are height-averaged and they also applied the Bernoulli
parameter to argue the presence of outflow. However, Abramowicz et
al.(2000) pointed out that the positive Bernoulli function (instead
of Bernoulli parameter, which is defined in convenience of
self-similar assumptions) is not sufficient for outflow (also see
the simulation works done by Stone et al. 1999 and Yuan \& Bu 2010).
Blandford \& Begelman (2004) furthered their work and presented some
self-similar two-dimensional solutions of radiatively inefficient
accretion flows with outflow. They assumed hydrostatic equilibrium
in the vertical direction and that convection dominates the heat
transport, which may only be applicable in certain cases. Xu \& Chen
(1997, hereafter XC97) relaxed $v_\theta=0$ and obtained two types
of solutions with outflow: accretion and ejection solutions. However
their solutions require the net accretion rate to be 0, which is not
realistic. Xue \& Wang(2005, hereafter XW05) followed NY95 and
solved the disk structure along the $\theta$ direction considering
$v_\theta$. They arbitrarily set a disk surface, at which $v_r=0$,
and the sound speed on the surface for their calculation. Their
solutions display a field of inflow near the equatorial plane with
wind blowing out of the upper boundary, however the boundary is set
as an input parameter rather than being calculated, and they only
investigated several cases of ADAFs. In \S{3.2} we will see that
according to our calculation, their assumption doesn't hold for
accretion flows with large $\alpha$ value. Sadowski et al.(2010)
abandoned the self-similar assumptions and solved the accretion disk
structure in the radial and vertical direction simultaneously.
Because the Navier-Stokes equations for accretion disks cannot be
decoupled intrinsically, they adopted other assumptions, e.g. the
disk is not geometrically thick, to derive the equations. As in
their work they didn't consider $v_z$ while $v_r$ and $v_\phi$ were
supposed not to vary vertically, they were not able to study
outflows. In summary, in the analytical model study, the vertical or
$\theta$-direction structure of accretion disks and outflows can
still not be dealt with satisfactorily, and many improvements are
still needed to be made.

On the other hand, observationally there are more and more evidences
for outflows in accretion systems, such as Sgr A*(Marrone et al.
2006; Xie \& Yuan 2008), soft X-ray transients(Loeb et al. 2001) and
quasars with blueshifted absorption lines(e.g., PG1115+80; Chartas
et al. 2003). Many numerical simulation works have also discovered
outflows in their results (e.g. Stone et al., 1999; Igumenshchev \&
Abramowicz, 2000; Okuda et al. 2005; Ohsuga et al.2005; Ohsuga \&
Mineshige 2007; Ohsuga et al. 2009) . The common existence of
outflows in these works inspire us to investigate the vertical
structure of accretion flows explicitly, to find solutions which can
deal with $v_\theta$ and positive $v_r$ to get a clear velocity
field, and with more reasonable boundary conditions. As a first
step, we followed the work done by NY95 and XW05, using self-similar
assumptions in the radial direction and solving the ODEs along the
$\theta$ direction in spherical  coordinates ($r\theta\phi$). We
used the $\alpha$ viscosity prescription and assumed that the
$r\phi$ component of the viscosity stress tensor is dominant. By
neglecting other components of the viscosity stress tensor, the
necessary number of boundary conditions is reduced and we will only
need the boundary conditions in the equatorial plane, which is
obatined by symmetry and is thus quite physical. As we didn't set
other arbitrary restrictions to $v_r$ and $v_\theta$ other than the
self-similar assumptions, we can get a velocity field containing
positive $v_r$, to discuss the flow structure with possible
outflows. These assumptions are applicable to different kinds of
accretion disk models, ranging through SSD, ADAF and slim disk, so
we also did many calculations with different sets of parameters
(which is very difficult to do in a numerical simulation work as it
would be too time-consuming), trying to find the flow structure
dependence on the parameters, in order to understand the
inflow/outflow mechanism more physically. These results can also be
helpful to future numerical simulation works when they set the input
parameters.

It should be noted that there is another branch of researches which
investigate accretion flows with an outflow (usually wind) plus
accretion disk model. In these researches, the configuration of the
outflow and accretion disk is usually preset, and the calculations
are focused on either the outflow or the accretion disk while the
influence of the other part is simplified or parameterized (e.g.
Fukue, 1989; Takahara et al., 1989; Kusunose, 1991; BB99; Misra \&
Taam, 2001; Fukue, 2004; Xie \& Yuan, 2008). Recently there are also
several works done in this way which deal with the outflow and the
accretion disk simultaneously (Kawabata \& Mineshige, 2009; Dotan \&
Shaviv, 2010). Compared with our work, in these studies the
accretion disk is usually height-integrated and the configuration of
the accretion flow is assumed rather than being calculated, while in
our work we solve the full hydrodynamic equations to get the
configuration of the accretion flow. Our work focuses on studying
the general structure of disks, outflows and the physical mechanism
behind them, and the results can be complementary to one another.

In \S{2} we present the basic equations and assumptions we used in
our calculations. In \S{3.1} we discuss our numerical methods and
present solutions corresponding to typical parameters of SSD, ADAF
and slim disk. In \S{3.2} we show the disk structure dependence on
different parameters and discuss their physical meanings. In \S{4}
we present our summary and discussion.

\section{EQUATIONS AND ASSUMPTIONS}

We consider the hydrodynamic equations of an accretion flow in the
spherical coordinates ($r\theta\phi$). The flow is assumed to be
steady($\partial /\partial t=0$) and axisymmetric($\partial
/\partial \phi=0$). Thus the continuity equation can be written as
\begin{equation}\label{1}
    \frac{1}{r^2}\frac{\partial}{\partial r}(r^2 \rho
    v_r)+\frac{1}{r \sin \theta}\frac{\partial}{\partial
    \theta}(\sin \theta \rho v_{\theta})=0 .
\end{equation}
We assume that for the accretion flow, only the $r\phi$-component of
the viscous tensor, $t_{r\phi}$, is dominant. And we use the
Newtonian gravitational potential, $\Phi=-GM/r$. Then the equations
of motion can be written as(Kato et al. 2008, also see Appendix A of
XW05)
\begin{equation}\label{2}
    v_r \frac{\partial v_r}{\partial
    r}+\frac{v_{\theta}}{r}(\frac{\partial v_r}{\partial
    \theta}-v_\theta)-\frac{v_\phi^2}{r}=-\frac{GM}{r^2}-\frac{1}{\rho} \frac{\partial
    p}{\partial r}
\end{equation}

\begin{equation}\label{3}
    v_r \frac{\partial v_\theta}{\partial
    r}+\frac{v_{\theta}}{r}(\frac{\partial v_\theta}{\partial
    \theta}+v_r)-\frac{v_\phi^2}{r} \cot \theta=-\frac{1}{\rho r} \frac{\partial
    p}{\partial \theta}
\end{equation}

\begin{equation}\label{4}
    v_r \frac{\partial v_\phi}{\partial
    r}+\frac{v_{\theta}}{r} \frac{\partial v_\phi}{\partial
    \theta}+\frac{v_{\phi}}{r}(v_r+v_\theta \cot \theta)=\frac{1}{\rho r^3} \frac{\partial
    }{\partial r}(r^3 t_{r\phi})
\end{equation}
In our calculation, we adopt the $\alpha$ prescription of viscosity
$t_{r\phi}=-\alpha p$ and $p$ is the total pressure. The shearing
box radiation Magnetohydrodynamics (MHD) simulations done by Hirose
et al. (2009) found that the vertically integrated stress is
approximately proportional to the vertically averaged total thermal
(gas plus radiation) pressure. In our work, however, we also apply
this relation locally.

Following NY95, we use the advective factor, $f\equiv
Q_\mathrm{adv}/Q_\mathrm{vis}$, i.e. a fraction $f$ of the
dissipated energy is advected as stored entropy and a fraction
$(1-f)$ is lost due to radiation. In our calculation we assume that
$f$ is constant in the accretion flow, so the energy equation can be
written as
\begin{equation}\label{5}
    \rho (v_r\frac{\partial e}{\partial r}+\frac{v_\theta}{r}\frac{\partial e}{\partial
    \theta})-\frac{p}{\rho}(v_r\frac{\partial \rho}{\partial
    r}+\frac{v_\theta}{r}\frac{\partial \rho}{\partial \theta})=f
    t_{r\phi} r \frac{\partial }{\partial r}(\frac{v_\phi}{r}) ,
\end{equation}
where $e$ is the internal energy of the material and can be
expressed as(Kato et al. 2008)
\begin{equation}\label{6}
    \rho e=\frac{p_\mathrm{gas}}{\gamma-1}+3p_\mathrm{rad},
\end{equation}
where $\gamma$ is the heat capacity ratio and is considered to be a
constant input parameter in our calculation.

We adopt the self-similar assumptions in the radial direction,
therefore we seek a solution of the form
\begin{eqnarray}
  \rho &=& \rho(\theta)r^{-n}, \\
  v_r &=& v_r(\theta)\sqrt{\frac{GM}{r}}, \\
  v_\theta &=& v_\theta(\theta)\sqrt{\frac{GM}{r}}, \\
  v_\phi &=& v_\phi(\theta)\sqrt{\frac{GM}{r}}, \\
  p &=& p(\theta)GMr^{-n-1}.
\end{eqnarray}
This set of self-similar solutions is similar to that of NY95 and is
actually identical to that of XW05. Compared with those works, here
we use the total pressure $p$ in Equation (11) instead of the sound
speed $c_s$, which is proportional to $\sqrt{p/\rho}$. As stated in
NY95, the only length scale in the problem is $r$ and the only
frequency is $\Omega_\mathrm{K}$, thus all velocities must scale
with radius as $r\Omega_\mathrm{K}$. In NY95, Narayan and Yi set
$n=3/2$, which implies $v_\theta=0$ according to the continuity
equation, thus they intrinsically set no outflow for the accretion
disk. Here we relax this parameter $n$ following BB99 and XW05, to
allow outflows from the disk.

If we substitute Equations (7)-(11) into Equations (1)-(4), the $r$
components can be eliminated, leaving only the dimensionless
fuctions $\rho(\theta)$, $v_r(\theta)$, $v_\theta(\theta)$,
$v_\phi(\theta)$ , $p(\theta)$, the variable $\theta$ and some
constants which are set as input parameters. The energy equation
(Equation (5)) is somewhat more complex. As the internal energy
depends differently on gas pressure and radiation pressure, we
discuss the energy equation more carefully here. First, one may
notice that only the total pressure $p$ appears in Equations
(1)-(4). This is not hard to understand as the dynamical processes
don't recognize whether the pressure is from gas or radiation.
$p_\mathrm{gas}$ and $p_\mathrm{rad}$ do affect the energy equation
in a different way, and to describe this effect, we define the gas
pressure ratio $\beta$,
\begin{equation}\label{7}
    \beta \equiv
\frac{p_\mathrm{gas}}{p}=\frac{p_\mathrm{gas}}{p_\mathrm{gas}+p_\mathrm{rad}}.
\end{equation}
There is one particular case that when $\gamma=4/3$, no matter what
the value of $\beta$ is, the solution remains the same, as in this
case Equation (6) becomes $\rho e=3p$; if the accretion flow is
radiation pressure dominated ($\beta \rightarrow 0$), then Equation
(6) also becomes $\rho e=3p$, and the flow structure won't depend on
$\gamma$; if the accretion flow is gas pressure dominated, then
Equation (6) becomes $\rho e=p/(\gamma-1)$, and the result will
depend on the value of $\gamma$. For a general case, Equation (6)
can be written as
\begin{equation}\label{8}
    \rho
    e=\frac{p_\mathrm{gas}}{\gamma-1}+3p_\mathrm{rad}=\left[\frac{\beta}{\gamma-1}+3(1-\beta)\right]p.
\end{equation}
Physically $\beta$ is a value between 0 and 1, so $\rho e$ is always
between $3p$ and $p/(\gamma-1)$. Then we can expect that a general
solution should be somewhere between two extreme cases: the gas
pressure dominated case and the radiation pressure dominated case.
For these two extreme cases, we can calculate the exact solutions of
the problem, as Equation (6) will be simplified to forms without
$\beta$. These two cases also provide the typical solutions we
emphasize in \S{3.1}. However we also want to know how the solutions
change from the gas pressure dominated case to the radiation
pressure dominated case. For the accretion flows in which both gas
pressure and radiation pressure are important, we have to assume the
gas pressure ratio $\beta$ to be constant to solve the problem.
Solving the problem with variant $\beta$ is beyond the capability of
this self-similar treatment, and we're planning to do that in our
next work.

As we already mentioned, the radiation pressure dominated case has
the same result as that of the gas pressure dominated case with
$\gamma=4/3$, so the question eventually becomes examining how the
solution changes according to $\gamma$ for the gas pressure
dominated case. We define an equivalent $\gamma$ here
\begin{equation}\label{9}
    \gamma_\mathrm{equ}\equiv \frac{p}{\rho
    e}+1=\frac{\gamma-1}{\beta+3(1-\beta)(\gamma-1)}+1,
\end{equation}
so that Equation (6) becomes
\begin{equation}
    \rho e=\frac{p}{\gamma_\mathrm{equ}-1},
\end{equation}
and this $\gamma_\mathrm{equ}$ represents the equivalent $\gamma$
that the accretion flow would have if the flow is treated as gas
pressure dominated (even $\beta\neq 1$). We can see that if the flow
is radiation pressure dominated ($\beta=0$), then no matter what the
value of $\gamma$ is, $\gamma_\mathrm{equ}$ is always 4/3. Here we
treat $\gamma$ as a constant, and for any specific case, we can get
a constant $\gamma_\mathrm{equ}$ and all the situations can be
treated as a gas pressure dominated flow with $\gamma_\mathrm{equ}$
as an input parameter. Usually $\gamma$ is taken as a value between
7/5(the value for diatomic ideal gas) and 5/3(the value for
monatomic ideal gas), so that $\gamma_\mathrm{equ}$ ranges between
4/3 and 5/3.

With Equations (7)-(11) and (15), the Equations (1)-(5) can be
reduced to a set of five ODEs:

\begin{equation}\label{10}
    2v_\theta(\theta)\frac{d\rho(\theta)}{d\theta}+\rho(\theta)\left[(3-2n)v_r(\theta)+2\left(\cot{\theta}v_\theta(\theta)+\frac{d
    v_\theta(\theta)}{d\theta}\right)\right]=0,
\end{equation}

\begin{equation}\label{11}
    2(n+1)p(\theta)+\rho(\theta)\left[v_r(\theta)^2+2\left(-1+v_\theta(\theta)^2+v_\phi(\theta)^2-v_\theta(\theta)\frac{dv_r(\theta)}{d\theta}\right)\right]=0,
\end{equation}

\begin{equation}\label{12}
    2\frac{dp(\theta)}{d\theta}+\rho(\theta)\left[-2\cot{\theta}v_\phi(\theta)^2+v_\theta(\theta)\left(v_r(\theta)+2\frac{dv_\theta(\theta)}{d\theta}\right)\right]=0,
\end{equation}

\begin{equation}\label{13}
    -2(n-2)\alpha
    p(\theta)+\rho(\theta)\left[v_r(\theta)v_\phi(\theta)+2v_\theta(\theta)\left(\cot{\theta}v_\phi(\theta)+\frac{dv_\phi(\theta)}{d\theta}\right)\right]=0,
\end{equation}

\begin{equation}\label{14}
    2\rho(\theta)v_\theta(\theta)\frac{dp(\theta)}{d\theta}+p(\theta)\left\{\rho(\theta)\left[2(n\gamma_\mathrm{equ}-n-1)v_r(\theta)-3\alpha
    f(\gamma_\mathrm{equ}-1)v_\phi(\theta)\right]-2\gamma_\mathrm{equ}
    v_\theta(\theta)\frac{d\rho(\theta)}{d\theta}\right\}=0,
\end{equation}
with 5 dimensionless fuctions $\rho(\theta)$, $v_r(\theta)$,
$v_\theta(\theta)$, $v_\phi(\theta)$ , $p(\theta)$, the variable
$\theta$ and four input parameters
($\alpha$,$f$,$\gamma_\mathrm{equ}$,$n$). This set of ODEs can be
numerically solved with proper boundary conditions. We assume the
structure of the disk is symmetric to the equatorial plane, and thus
we have
\begin{equation}\label{15}
    \theta=90\textordmasculine:
    v_\theta=\frac{d\rho}{d\theta}=\frac{dp}{d\theta}=\frac{dv_r}{d\theta}=\frac{dv_\phi}{d\theta}=0
\end{equation}
in which only four conditions are independent. For the last boundary
condition we set $\rho(90\textordmasculine)=1$, which can be
normalized by a scale factor if the effective accretion rate at a
certain radius is set (NY95, XW05, etc.). These boundary conditions
are enough for our calculations and we didn't introduce other
arbitrary boundary conditions.

\section{NUMERICAL RESULTS}

\subsection{Typical Solutions}

We obtained numerical solutions of Equations (16)-(20) with
different sets of input parameters
($\alpha$,$f$,$\gamma_\mathrm{equ}$,$n$). Some typical solutions are
shown in Figures 1-4. The calculation starts from the equatorial
plane ($\theta=90\textordmasculine$) towards the vertical axis
($\theta=0\textordmasculine$). Generally the mass density $\rho$ and
total pressure $p$ will decrease as $\theta$ decreases, and at
certain inclination they will get very close to 0 as shown in the
figures. We take this as the upper boundary of the flow structure.
If we continue the calculation through this inclination, it will
encounter numerical errors. We think the reason is that we can't
describe the flow near the vertical axis with a simple self-similar
solution in the radial direction. This effect is also to be
expected. If we describe the upper boundary (minimum $\theta$) that
we reach in our calculation as $\theta_\mathrm{b}$, then the
effective accretion rate $\dot{M}_\mathrm{eff}$ across a sphere at
radius $r$ within the region calculated by us is
\begin{equation}\label{16}
    \dot{M}_\mathrm{eff}=2\int_{\theta_\mathrm{b}}^{90\textordmasculine}\rho v_r \cdot 2\pi r
    \sin{\theta} r \cdot (\pi/180\textordmasculine)d\theta=4\pi\sqrt{GM}
    r^{\frac{3}{2}-n}\int_{\theta_\mathrm{b}}^{90\textordmasculine}
    v_r(\theta)\rho(\theta)\sin{\theta}\cdot (\pi/180\textordmasculine)d\theta,
\end{equation}
which is a function of $r$ unless $n=3/2$. In Equation (22) (and
subsequently in this paper) negative values of
$\dot{M}_\mathrm{eff}$ represent inflow while positive values
represent outflow. If we describe the accretion rate in the region
between the vertical axis and the inclination $\theta_\mathrm{b}$ as
$\dot{M}_\mathrm{axis}$, then according to the steady nature of the
flow, we have
\begin{equation}\label{17}
    \dot{M}_\mathrm{eff}+\dot{M}_\mathrm{axis}=\dot{M}
\end{equation}
in which $\dot{M}$ represents the total accretion rate across any
sphere at a reasonable radius centered by the central accretor and
should be a constant for a steady accretion flow. If the solution
doesn't end at an upper boundary, and instead can describe the
entire flow structure in the whole space, then $\theta_\mathrm{b}=0$
and $\dot{M}_\mathrm{eff}=\dot{M}$, which should be a constant.
According to Equation (22), this can only happen in the following
two cases: (1) $n=3/2$ which enforces $\dot{M}_\mathrm{eff}$ not to
change with radius $r$; (2) when $n\neq 3/2$, the integration term
in Equation (22) must be 0, in which
    case $\dot{M}_\mathrm{eff}=0$ and is a constant along radius
    $r$. The first case was discussed in NY95. Because $n=3/2$,
    $r^2\rho v_r$ is independent of $r$, and the continuity equation
    (1) shows that $v_\theta=0$, resulting in a solution in which the flow
is always radial (with rotation). The second case was discussed
    in XC97, and the fact that $\dot{M}=0$ requires that the outflow
    rate exactly equals the inflow rate at any radius. However this is
    unrealistic for an accretion flow, which is also discussed in
    XW05. The reason is that, when material is accreted in the form
    of an accretion flow, gravitational energy is released and part
    of it is changed to internal energy via viscous friction. The
    restriction that outflow rate equals inflow rate requires that the internal
    energy released from gravitational energy should
be fully returned to gravitational energy, which violates the second
law of thermodynamics. Therefore, the inflow rate must be larger
than the outflow rate at a certain radius, in order to compensate
for the increase in entropy in the hydrodynamic process. As we have
mentioned before, there are many observations of accretion systems
with outflows, and they can be neither of the two cases mentioned
above (case (1) has no outflow while case (2) violates the second
law of thermodynamics). So the self-similar solutions for an
accretion disk with outflow have to be truncated at some
inclination. It should be noted that this effect has already been
discussed in XW05, though they assumed that the self-similar
solution is only valid for the inflow part. Our solutions, however,
show that at least part of the outflow (with
$\theta_\mathrm{b}<\theta<\theta_0$) can be described with
self-similar assumptions.

From the above analysis, we can also see that $n$ is a very
important parameter. When $n=3/2$, the entire flow structure can be
described by the same set of self-similar solutions, however this
kind of solutions have no outflow. When $n \neq 3/2$, self-similar
solutions can only describe part of the entire flow structure.
According to Equation (22), when $n<3/2$, the effective accretion
rate $\dot{M}_\mathrm{eff}$ decreases as $r$ decreases, implying
that material is lost due to outflows. This kind of solutions
contain outflows, which is also consistent with the results of many
numerical simulation works (e.g. Okuda et al. 2005; Ohsuga et
al.2005; Ohsuga \& Mineshige 2007; Ohsuga et al. 2009). On the other
hand, when $n>3/2$, the solutions have an effective accretion rate
$\dot{M}_\mathrm{eff}$ that increases towards the central accretor,
which implies that there must be matter injection into the accretion
flow from high latitudes. This kind of matter injection is not
likely to happen in real cases. So we would like to investigate
solutions with $n\leqslant 3/2$. Solutions with $n=3/2$ have been
examined in NY95, and they are only for ADAFs, while solutions with
$n<3/2$ in our work is applicable to a much wider range of accretion
disk types. So in this subsection, we take $n=1.3$ to calculate the
typical solutions. In this way we can observe how the solutions
differ from those of $n=3/2$ when $n$ doesn't change much. We also
calculated how the solutions change with different $n$, which will
be discussed in detail in \S{3.2}. The result shows that, when other
parameters are the same, for a large range of $n$ with $n<3/2$ the
solutions display similar structures qualitatively.

The value of the viscosity parameter $\alpha$ can be inferred both
from observations and from MHD simulations of magneto-rotational
instability (MRI). As reviewed by King et al. (2007), the value of
$\alpha$ for ionized disks given by numerical simulations is $\sim
0.01$, while that obtained through observations is $\sim 0.1 - 0.4$.
However, the observational determinations of $\alpha$ are strongly
model-dependent and it still remains an open question what the value
of $\alpha$ should be for hot, ionized disks. In this subsection we
take $\alpha=0.1$, which is a typical value used in theoretical
models of accretion disks. We also discussed how the solutions would
change with different $\alpha$ in \S{3.2}.

Among the most popular analytical disk models are the SSD, ADAF and
slim disk models, so here we show the solutions with input parameter
corresponding to them respectively. For accretion flows composed
mostly of fully ionized hydrogen, the material can be regarded as
monatomic ideal gas, of which $\gamma=5/3$, and that is the value we
take here (accretion flows composed of unionized $\mathrm{H}_2$
would have $\gamma=7/5$. We didn't discuss this situation here, but
some relative results can be found in \S{3.2}). Figures 1 \& 2 are
both for the SSD model which has little advection ($f=0.01$), while
Figure 1 is for the gas pressure dominated region and Figure 2 for
the radiation pressure dominated region. The ADAF model is treated
as advection dominated($f=1$), and gas pressure dominated
($\beta=1$, so that $\gamma_\mathrm{equ}=5/3$), as shown in Figure
3. The slim disk model is treated as advection dominated($f=1$), and
radiation pressure dominated ($\beta=0$, so that
$\gamma_\mathrm{equ}=4/3$), as shown in Figure 4. The other two
parameters are all set as $\alpha=0.1$ and $n=1.3$. In each figure,
we show the function curves of $v_r(\theta)$, $v_\theta(\theta)$,
$v_\phi(\theta)$ , $p(\theta)$, $\rho(\theta)$ and the Mach numbers
of $v_r$ and $v_\theta$, in which the adiabatic sound speed is
calculated as
\begin{equation}\label{c_s}
    c_s=\sqrt{\gamma\frac{p}{\rho}}.
\end{equation}
As shown in Equations (7-11), all the velocities scale with
$\sqrt{GM/r}$. The density $\rho(\theta)$ is scaled to 1 in the
equatorial plane and it can be determined in a real case once the
accretion rate at a certain radius is given. The pressure $p$ is
also scaled, and it satisfies the relation that
$p(\theta)/\rho(\theta)=(p/\rho)/(GM/r)$. So once we get the actual
density $\rho$ at a certain radius, we can calculate the
corresponding pressure $p$ easily. The Mach number curves for $v_r$
first decreases with $\theta$, reaching 0 at some inclination, and
then increases. That is because it changes sign along $\theta$ and
we use absolute values in calculating the Mach numbers.

There are several common features among Figures 1-4. The value of
$v_\theta$ is always non-positive. The value of $v_r$ is always
negative close to the equatorial plane, and becomes positive at
smaller inclinations. In each of these figures, there exists a
certain inclination $\theta_0$ at which $v_r(\theta)=0$. This
$\theta_0$ has similar meaning with the '$\theta_0$' in XW05, except
that it is obtained through calculation rather than being preset as
an input parameter in XW05. In the region
$\theta_0\leqslant\theta\leqslant 90\textordmasculine$,
$v_r(\theta)$ is negative, meaning that the accretion flow is moving
towards the central accretor as inflow. So this region corresponds
to the 'normal' accretion disk part in basic models and we call it
the inflow region in this paper. In the region
$\theta_\mathrm{b}<\theta< \theta_0$, $v_r(\theta)$ is positive and
the accretion flow is moving away from the central accretor, and we
call this region the outflow region. The region between the upper
boundary and the vertical axis contains the outflow which blows out
of the upper boundary in the form of wind as shown in Figure 5, and
we call this region the wind region, to distinguish it from the
outflow region. It should be noted that our solutions actually
cannot describe the structure of the wind region. However, it is
natural to suppose that the flow structure in the wind region with
$0\leqslant \theta<\theta_\mathrm{b}$ is also outflow, and the
outflow region ($\theta_\mathrm{b}<\theta< \theta_0$) can either be
regarded as a transition region between inflow and outflow, or as
part of the outflow which can still be described with the
self-similar assumptions.

The physical properties shown in Figures 1 \& 2 agree quite well
with the SSD model. The rotation is very close to being Keplerian;
the radial velocity is much smaller than the azimuthal velocity; and
the disk is geometrically thin. However, as mentioned above, our
calculation shows that even in the SSD case, there exists outflow in
the accretion disk. As shown in Figures 5(a) \& 5(b), this outflow
is in the form of wind blowing out from the disk surface. Both $v_r$
and $v_\theta$ are subsonic in the SSD cases.

The advection dominated solutions, as shown in Figures 3 \& 4, are
geometrically much thicker than the solutions with low advection,
and are able to describe the structure of the majority of the space,
except only a small region near the vertical axis. The rotation is
sub-Keplerian and the radial velocity can be comparable with the
azimuthal velocity. Both the ADAF and the slim disk case show strong
outflows compared with the SSD cases, which can be seen more clearly
in Figure 5. $v_\theta$ keeps to be subsonic for both cases.
However, for the slim disk case (Figure 4), $v_r$ becomes supersonic
near the vertical axis, displaying a stronger outflow than the other
three typical solutions.

Outflow is the result of the competition among the pressure
gradient, the centrifugal force and the gravitational force. As we
have mentioned before, the key feature of the outflow in our work is
positive $v_r$, so we would like to investigate the properties of
these forces along the radial direction. The radial components of
the pressure gradient, the centrifugal force and the gravitational
force can be written respectively as
\begin{eqnarray}
  -\frac{1}{\rho}\frac{\partial p}{\partial r} &=& A(\theta)\frac{GM}{r^2}, \\
  \frac{v_\phi^2}{r} &=& B(\theta)\frac{GM}{r^2}, \\
  \frac{GM}{r^2} &=& 1\cdot\frac{GM}{r^2},
\end{eqnarray}
in which
\begin{eqnarray}
  A(\theta) &=& (1+n)\frac{p(\theta)}{\rho(\theta)} \\
  B(\theta) &=& v_\phi^2(\theta).
\end{eqnarray}
 We plot the values of $A(\theta)$, $B(\theta)$ and $A(\theta)+B(\theta)$ for the four typical solutions in Figure 6.
The dotted lines correspond to the gravitational force which is
scaled as 1; the dashed lines correspond to the radial component of
the centrifugal force; the dot-dashed lines correspond to the radial
component of the pressure gradient; and the solid lines correspond
to the sum of the radial components of the centrifugal force and the
pressure gradient, which drives the outflow. Figure 6(a) shows the
properties of gas pressure dominated SSDs, while Figure 6(b) shows
the properties of radiation pressure dominated SSDs. It can be seen
that for both of the SSD cases, the influence of pressure gradient
is very small, and almost in the entire range of the solution, the
gravitational force is balanced with the sum of the centrifugal
force and the pressure gradient in the radial direction. This is
consistent with the SSD model in which both the advection and the
pressure gradient are small enough to be neglected. Therefore these
two solutions have similar accretion flow structure and only some
quantitative differences exist between each other.

 On the other hand, for the
advection dominated cases, such as the ADAF case (Figure 6(c)) and
the slim disk case (Figure 6(d)), the pressure gradient plays a more
important role than the centrifugal force and drives a significant
outflow as $\theta$ decreases. The slim disk case has qualitative
differences from the other three cases. In both SSD cases and the
ADAF case, the radial component of pressure gradient starts
decreasing at smaller inclination angle, and to balance this effect,
the disk rotates faster as $\theta$ decreases to increase the
centrifugal force, as shown in Figures 1-3. However in the slim disk
case, the radial component of the pressure gradient keeps increasing
as $\theta$ decreases. It not only drives a much stronger outflow
than the other three cases, but also reduces the required
centripetal force to keep the disk rotating. Thus the disk rotates
slower as $\theta$ decreases, as shown in Figure 4.

As a comparison with NY95, we also present here a solution with
parameters $\alpha=0.1$, $n=3/2$, $\gamma_\mathrm{equ}=1.6061$ and
$f=1$ in Figure 7. This set of parameters corresponds to
'$\epsilon\prime=0.1$' in NY95. In this case, the accretion flow
doesn't contain outflow and the three components of velocity and the
sound speed remain constant in the structure. The curves of
$\rho(\theta)$ and $v_r(\theta)$ don't agree well with NY95, however
this is because NY95 also sets the boundary conditions on the
vertical axis and the values of $\rho(\theta)$ and $v_r(\theta)$ are
limited by these boundary conditions. Our solution only uses the
boundary conditions in the equatorial plane, and can still reach the
vertical axis in this case and has a reasonable distribution of
physical properties, which further proves the applicability of our
method.

\subsection{Solution Dependence on Input Parameters}

In this section we show how the solutions change according to
different input parameters respectively, and discuss their physical
meanings. As shown in \S{3.1}, the whole space is generally divided
into three regions: an inflow region, an outflow region and a wind
region. As the accretion flow is symmetrical both axially and under
reflection in the equatorial plane, we only need to investigate the
flow structure in a quadrant of the $r\theta$ plane. In this case,
the inflow region starts from the equatorial plane
($\theta=90\textordmasculine$) to the inclination $\theta=\theta_0$
where $v_r(\theta)=0$; the outflow region starts from the
inclination $\theta=\theta_0$ to the upper boundary of the flow
structure at $\theta=\theta_\mathrm{b}$, and beyond that is the wind
region ($0<\theta<\theta_\mathrm{b}$) in which the motion of the
matter doesn't satisfy the self-similar assumptions. Here we use
$\theta_0$ and $\theta_\mathrm{b}$ as indicators of the disk size in
$\theta$-direction, which directly reflects the size of the inflow
region and the whole inflow/outflow region, respectively. In the
figures showing how the properties of solutions change according to
input parameters, each line represents a set of solutions, and we
assign a unique label for each solution set for clarity. Table 1
lists all the solution set labels, their input parameters and
corresponding lines at the end of \S{3.2}.

\texttt{\textbf{1) Solution dependence on $\alpha$}}

The dependence of $\theta_0$ and $\theta_\mathrm{b}$ on $\alpha$ is
shown in Figure 8. First we can clearly see that over a large range
of $\alpha$, the advection dominated solutions have both larger
inflow region and larger inflow/outflow region. That's because in
these solutions most of the viscous heating is stored in the
accretion flow as internal energy, resulting in larger pressure at
given density and thus larger size of the whole inflow/outflow
region.

For solutions with the same advective factor $f$, gas pressure
dominated solutions generally have larger inflow region. The inflow
region lasts from the equatorial plane to the inclination $\theta_0$
where $v_r$ reaches 0, so its size depends on the behavior of $v_r$.
In the equatorial plane, the reflection symmetry gives Equation
(21), and when it is substituted into Equations (17), (19) \& (20),
we can get
\begin{eqnarray}
  2(n+1)\frac{p(\theta)}{\rho(\theta)}+v_r^2(\theta)+2(v_\phi^2(\theta)-1) &=& 0, \\
  -2(n-2)\alpha\frac{p(\theta)}{\rho(\theta)}+v_r(\theta)v_\phi(\theta) &=& 0, \\
  2(n\gamma_\mathrm{equ}-n-1)v_r(\theta)-3\alpha f(\gamma_\mathrm{equ}-1)v_\phi(\theta) &=& 0.
\end{eqnarray}
Equation (30) describes the hydrodynamical balance in the radial
direction among the pressure gradient, the centrifugal force and the
gravitational force, which determines the acceleration of $v_r$ (the
term $v_r^2(\theta)$ represents the acceleration of $v_r$ under the
self-similar assumptions). Equation (31) describes how the viscosity
affects the angular momentum transfer rate in the equatorial plane.
Equation (32) describes the energy mechanism of the accretion flow
that a fraction $f$ of the dissipated energy is advected as stored
entropy. It should be noted that $v_r$ is generally negative in the
equatorial plane, so the absolute value of $v_r$ can be written as
(according to Equation (32))
\begin{equation}\label{|v_r|}
    |v_r(90\textordmasculine)|=-v_r(90\textordmasculine)=\frac{3\alpha
    f}{2n}v_\phi(90\textordmasculine)\left[\frac{1}{1-n(\gamma_\mathrm{equ}-1)}-1\right].
\end{equation}
If $v_\phi(90\textordmasculine)$ remains the same, then in the range
of parameters we study in this paper, $|v_r(90\textordmasculine)|$
increases with $\gamma_\mathrm{equ}$. However
$v_\phi(90\textordmasculine)$ also changes with input parameters,
and the exact value of $v_r(90\textordmasculine)$ should be
calculated via Equations (30)-(32). Figure 10 shows the profile of
$v_r(90\textordmasculine)$ vs $\gamma_\mathrm{equ}$ when other
parameters are taken as $n=1.3$ and $\alpha=0.1$. It can be seen
that for both the advection dominated solutions and the solutions
with little advection, $|v_r(90\textordmasculine)|$ increases with
$\gamma_\mathrm{equ}$. With a larger value of $v_r$ in the
equatorial plane, the accretion flow is inclined to undergo a larger
range of $\theta$ to change $v_r$ to 0, resulting in a larger inflow
region. For the two SSD cases, as their flow structures are quite
similar, larger values of $v_r$ in the equatorial plane give the gas
pressure dominated solution set $a_1$ larger inflow regions than
$a_2$. For the advection-dominated cases, not only the equatorial
value of $v_r$ is smaller in the slim disk case than that in the
ADAF case, but its radial direction pressure gradient also increases
as $\theta$ decreases, giving a larger outward acceleration to
$v_r$, thus the gas pressure dominated solutions $a_3$ have larger
inflow regions than the radiation pressure dominated solutions
$a_4$.

The solution set $a_4$, which corresponds to the slim disk model,
has a large outflow size over a broad range of $\alpha$, displaying
a large outflow area. The solutions corresponding to the SSD model,
$a_1$ and $a_2$, have very small outflow regions, and their outflows
are mainly in the form of wind blowing out of the disk surface
instead of escaping in the outflow region. The solution set $a_3$,
which corresponds to the ADAF model, has a moderate outflow region
with small $\alpha$. However as $\alpha$ increases, this outflow
region quickly disappears and the outflow takes the form of wind
blowing out of the upper surface. The velocity field displayed in
Figure 5 is an example for $\alpha=0.1$.

$\alpha$ is connected with both the viscous heating and the angular
momentum transport of the accretion flow. Larger $\alpha$ will
increase the viscous heating for a fixed surface density. But at the
same time it also increases the angular momentum transfer rate, thus
increasing $v_r$ and decreasing surface density, which decreases the
total viscous heating rate. Therefore the dependence of the viscous
heating rate on $\alpha$, is somewhat complex. Viscous heating
represents how much gravitational energy is converted to internal
energy. But some of it is lost due to radiation and eventually there
is only an $f$ fraction of the viscous heating which is converted to
internal energy. The increase in the internal energy will raise the
temperature and subsequently the pressure, thus increasing the size
of the accretion flow. However, for the SSD cases shown in $a_1$ \&
$a_2$, $f$ is very small, thus the energy converted to internal
energy is negligible and the disk structure is dominated dynamically
by the gravity and the rotation of the material, rather than the
pressure gradient. As a result, the sizes of accretion flows change
very little with different $\alpha$ for $a_1$ \& $a_2$.

On the other hand, for the advection dominated solutions $a_3$ \&
$a_4$, a large fraction of the gravitational energy is eventually
stored as internal energy, so that the pressure gradient plays a
more important role in determining the flow structure as mentioned
in \S{3.1}. Therefore the sizes of accretion flows are closely
connected with $\alpha$. In Figure 8 we can see that for $a_3$ \&
$a_4$, when $\alpha$ is very small, the viscous process is not quite
effective, therefore both the inflow region size and the outflow
region size are small (it's easy to understand if we consider the
situation $\alpha=0$, in which case no accretion flow will be
formed). From this small value, along with the increasing of
$\alpha$, the viscous process starts to take effect and the whole
inflow/outflow region size starts to increase, until it reaches a
maximum size at $\alpha \sim 0.05$ for line $a_3^\prime$ and $\alpha
\sim 0.12$ for line $a_4^\prime$(the exact value of $\alpha$ for
maximum size also depends on other parameters). From then on, the
larger values of $v_r$ starts to dominate the total effect and the
corresponding accretion flow size starts to decrease. This property
can also be seen in the top left panel of Figure 17, which displays
the ratio of the outflow rate to the inflow rate.

It should be noted that, in Figure 8, the size change based on
$\alpha$ is different between the whole inflow/outflow region and
the pure inflow region. The size of the whole inflow/outflow region
decreases after a peak at a certain $\alpha$ while the size of the
pure inflow region keeps increasing. For solution sets $a_1$, $a_3$
and $a_4$, this difference is enough to generate a critical
$\alpha_\mathrm{c}$ respectively, at which the outflow region
totally disappears. So for $\alpha
> \alpha_\mathrm{c}$, the accretion flow that we can resolve
with self-similar assumptions is composed of pure inflow and we
didn't find a surface at which $v_r(\theta_0)=0$. That's why we say
that the assumption used in XW05 may not be true for large $\alpha$.
Kluzniak and Kita (2000) also found this critical $\alpha$ which was
calculated to be $\sqrt{15/32}\approx 0.685$ by them. In our work,
however, $\alpha_\mathrm{c}$ depends on other parameters and can
change significantly, as shown in Figure 8 ($\sim 0.96$ for $a_1$,
$\sim 0.24$ for $a_3$ and $\sim 0.74$ for $a_4$). For $a_2$, this
$\alpha_\mathrm{c}$ is above 1 and it's not displayed.

Figures 9 displays the $v_r$ distribution along the
$\theta$-direction for solutions with different $\alpha$. Figures
(a), (b), (c) and (d) correspond to the four sets of solutions in
Figure 8 respectively, and the solid, dashed and dotted lines
correspond to different $\alpha$ as indicated in the legends. The
range of the axes are adjusted to show more details in comparison.
We can see that in the solutions corresponding to the SSD model,
i.e. Figures 9(a) \& 9(b), $v_r$ is almost proportional to $\alpha$.
That is because when alpha increases, so does the angular momentum
transfer rate. For the SSD case, the accretion flow structure is
mostly determined by gravity and rotation as discussed in \S{3.1},
and the rotation is quasi-Keplerian, so the increase in angular
momentum transfer rate increases $v_r$ almost proportionally. A
special case in the equatorial plane is shown by Equation (31), in
which when $v_\phi$ is Keplerian, the value of $v_r$ is proportional
to $\alpha$. For the ADAF and the slim disk cases, because pressure
gradient plays an important role in the accretion flow structure,
$v_r$ is generally not proportional to $\alpha$, but still
positively correlated with $\alpha$ as shown in Figure 9(c) \& 9(d).
We can also see that some lines in Figure 9 don't contain positive
$v_r$. That's because the $\alpha$ values for these solutions are
greater than the critical value $\alpha_\mathrm{c}$ mentioned above.

\texttt{\textbf{2) Solution dependence on $n$}}

The parameter $n$ describes how the density changes along the radius
in our self-similar assumptions. As discussed in \S{3.1}, when
$n=3/2$, the entire flow structure can be described by the same set
of self-similar solutions, but doesn't contain outflow; when
$n<3/2$, the solution contains outflow but the self-similar
assumptions cannot be applied to the region near the vertical axis;
solutions with $n>3/2$ are unlikely to happen in real cases.

The dependence of $\theta_0$ and $\theta_\mathrm{b}$ on $n$ is shown
in Figure 11. Figures 12 \& 13 display the $v_r$ and $v_\theta$
distributions along the $\theta$-direction for solutions with
different $n$. We can see that solutions all achieve a maximum size
at $n=3/2$ for both the whole inflow/outflow region and the inflow
only region. This is because when $n=3/2$ the entire space can be
described by the same set of self-similar equations as discussed in
\S{3.1}. In this case, $v_\theta=0$ and mass is conserved on the
streamlines pointing exactly at the central accretor, thus no
outflow is required. Before reaching the maximum value at $n=3/2$,
the flow size is positively correlated with $n$ as a general trend.
As $n$ decreases, according to Equation (22) $\dot{M}_\mathrm{eff}$
decreases faster inwards, which means a larger mass loss rate into
the area beyond $\theta_\mathrm{b}$, resulting in the increase in
the size of the wind region and the decrease in the size of the
whole inflow/outflow region in Figure 11, and the increase in
$v_\theta$ in Figure 13.

In Figure 11 line $b_4^\prime$ has 2 peaks. According to Equation
(25), the radial pressure gradient is proportional to $(1+n)p/\rho$,
so when $p/\rho$ doesn't change much, the radial pressure gradient
increases with $n$. In advection-dominated solutions $b_3$ \& $b_4$,
the radial pressure gradient is more important than centrifugal
force in accelerating $v_r$, so for larger $n$ which is inclined to
have larger radial pressure gradient, one would expect to see faster
increase in $v_r$ and stronger outflow in the outflow region. These
features of $v_r$ can be seen in Figure 12. The radial pressure
gradient is dominant in the slim disk case as mentioned in \S{3.1},
so the acceleration of $v_r$ increases significantly with increasing
$n$ in the slim disk case, which can be seen in Figure 12(d) with a
much steeper profile of $v_r$ when $n=1.3$. This increased outflow
velocity in the outflow region can cause a steeper density drop
along $\theta$ direction towards the upper boundary and therefore
make the outflow region become smaller, creating a valley in line
$b_4^\prime$ around $n=1.4$.

The top right panel in Figure 17 shows how the ratio of the outflow
rate to the inflow rate changes depending on $n$. We can see that
when $n$ becomes small enough, the ratio changes very little with
$n$. In this range of $n$, the size of different regions also
doesn't change much in Figure 10, and the flow structure is similar
to each other.

For $\gamma_\mathrm{equ}=5/3$, we cannot get a reasonable set of
equatorial values for physical properties when $n$ is larger than
3/2. For $\gamma_\mathrm{equ}=4/3$, i.e. solutions $b_2$ and $b_4$,
we can still get reasonable values of equatorial physical properties
when $n$ is larger than 3/2, so that we can calculate solutions with
$n>3/2$ in this case. Solutions with $n>3/2$ have an effective
accretion rate $\dot{M}_\mathrm{eff}$ that increases towards the
central accretor, which implies that there must be matter injection
into the accretion flow from high latitudes. This kind of matter
injection is not likely to happen in real cases, so we won't pay
more attention to the solutions with $n>3/2$ here. Solutions with
$n=3/2$ and $n<3/2$ both exist and have quite different structure.
However, as in NY95, solutions with $n=3/2$ are only for ADAFs,
while solutions with $n<3/2$ in our work is applicable to a much
wider range of accretion disks. In this sense, we believe that
outflows should be common in various accretion disks.

\texttt{\textbf{3) Solution dependence on $f$}}

The dependence of $\theta_0$ and $\theta_\mathrm{b}$ on $f$ is shown
in Figure 14. It has already been discussed that as the advective
factor $f$ increases, the size of the whole inflow/outflow region
also increases, which is shown more clearly here in Figure 14. The
gas pressure dominated solutions have larger inflow region in
agreement with Figures 8 \& 11, and the reason has been discussed in
\S{3.2.1}.

As $f$ increases, the size of the outflow region increases faster
for radiation pressure dominated flows, displaying a larger outflow
size for solutions corresponding to the slim disk model. The reason
is that, as $f$ increases, a larger fraction of viscous heating is
converted to internal energy instead of being lost via radiation,
raising the temperature of the accretion flow. The gas pressure is
proportional to $T$ while the radiation pressure is proportional to
$T^4$, so in the radiation pressure dominated solutions, $p$
increases faster as $f$ increases. The increased pressure inflates
the outflow region and causes a stronger outflow. On the other hand,
as $f$ gets close to 1, eventually the outflow velocity at small
inclination becomes so large, that it will instead reduce the size
of the outflow region because the material lost in the outflow
region is limited. As shown in the bottom left panel of Figure 17,
at a certain $f\sim 0.9$, the fraction of material lost in the
outflow region reaches an upper limit, and with larger $f$, this
fraction almost remains unchanged. In this range of $f$, the larger
outflow velocity will instead shrink the size of the outflow region.
This causes line $c_2^\prime$ in Figure 14 to reach a peak of
$\sim90\textordmasculine$ at $f\sim 0.9$ (corresponding to Figure
17(c)), and to start decreasing as $f$ becomes larger.

\texttt{\textbf{4) Solution dependence on $\gamma_\mathrm{equ}$ (and
consequently $\beta$)}}

The dependence of $\theta_0$ and $\theta_\mathrm{b}$ on
$\gamma_\mathrm{equ}$ is shown in Figure 15. The top axis displays
the corresponding values of $\beta$ when $\gamma$ takes the value of
5/3. If the intrinsic heat capacity ratio is not 5/3 (e.g. when
dealing with proto-stellar accretion disks with unionized gas), the
corresponding $\beta$ in Figure 15 may not be correct, but one can
still calculate the correct $\beta$ with $\gamma_\mathrm{equ}$ and
$\gamma$ according to Equation (14). Figure 16 displays the $v_r$
and $v_\theta$ distributions along the $\theta$-direction for
solutions with different $\gamma_\mathrm{equ}$. The bottom right
panel of Figure 17 shows how the ratio of the outflow rate to the
inflow rate changes depending on $\gamma_\mathrm{equ}$ (and
consequently $\beta$).

In Figure 15, the size of the pure inflow region generally increases
with $\gamma_\mathrm{equ}$ (and consequently $\beta$). As shown in
Figure 10 \& 16 and discussed in \S{3.2.1}, solutions with larger
$\gamma_\mathrm{equ}$ (and consequently larger gas pressure ratio
$\beta$) have larger values of $v_r$ in the equatorial plane. This
effect combined with the acceleration properties of $v_r$, causes
solutions with larger $\gamma_\mathrm{equ}$ (and consequently larger
$\beta$) to have larger inflow region. For solutions with low
advection, as the outflow region is very small, the size of the
whole inflow/outflow region also displays similar profile with
increasing $\gamma_\mathrm{equ}$ (and consequently $\beta$). That's
why lines $d_1$, $d_1^\prime$ and $d_2$ all have positive slope in
Figure 15.

In Figure 17(d), there exists a transition point for the advection
dominated solutions, and for solutions with
$\gamma_\mathrm{equ}\lesssim 3/2$ (corresponding to $\beta \lesssim
2/3$), the ratio of the outflow rate to the inflow rate keeps being
large. As discussed before, for the advection dominated cases,
radiation pressure dominated solutions have larger pressure
gradient, shown as steeper velocity profiles in Figure 16, which
drives stronger outflow. As shown in Figure 17(d), the outflow to
inflow ratio increases as $\gamma_\mathrm{equ}$ decreases from 5/3
(which indicates the decreasing of gas pressure ratio $\beta$ from
1). However, the outflow rate cannot exceed the inflow rate and
eventually the increase will cease, reaching a transition point at
$\gamma_\mathrm{equ}\sim 3/2$ (corresponding to $\beta \sim 2/3$).
For $\gamma_\mathrm{equ}\lesssim 3/2$ (corresponding to $\beta
\lesssim 2/3$), the outflow to inflow ratio remains almost
unchanged. Further decrease in $\gamma_\mathrm{equ}$ will continue
to increase the outflow velocity, while the fraction of the inflow
lost in the outflow region is almost unchanged, so the size of the
outflow region will shrink. This feature can be also seen in Figure
15, which shows that the size of the outflow region for advection
dominated solutions first increases with decreasing
$\gamma_\mathrm{equ}$, reaching an upper limit at
$\gamma_\mathrm{equ}\sim 3/2$ (corresponding to $\beta \sim 2/3$),
and decreases with decreasing $\gamma_\mathrm{equ}$ after that. It
should be noted that, this transition value of $\gamma_\mathrm{equ}$
depends on other input parameters, and this value of $\sim 3/2$
corresponds to $\alpha=0.1$, $n=1.3$ and $f=1$. For the solutions
with low advection, the ratio of the outflow rate to the inflow rate
don't change much and are always very small as shown in Figure
17(d).

\texttt{\textbf{5) Bends in solutions}}

We have shown the solution dependence on different input parameters.
The solutions sometimes show bends which we would like to discuss
more specifically here.

The bend about $\alpha$ is closely connected with the energy
mechanism. As discussed in \S{3.2.1}, the dependence of viscous
heating on $\alpha$ is somewhat complex. It has a peak at a certain
value of $\alpha$ (which depends on other input parameters), and
decreases on both sides. For the advection dominated solutions $a_3$
\& $a_4$, most of the viscous heating is stored as internal energy
in the accretion flow, which raises the temperature and subsequently
the pressure, thus increasing the size of the accretion flow. On the
other hand, the increase in $\alpha$ also increases the absolute
value of $v_r$ of the inflow near the equatorial plane, in which
case more work needs to be done to drive an outflow, decreasing the
outflow region as $\alpha$ increases. So naturally the size of the
whole inflow/outflow region also has a peak at some value of
$\alpha$. As $\alpha$ increases from a very small value, at first
the increase in viscous heating dominates the total effect, and the
size of the accretion flow increases, until it reaches a maximum
value; from then on, the increase in $v_r$ starts to dominate the
total effect, which makes the outflow more difficult to produce and
also reduces viscous heating when $\alpha$ becomes large enough,
thus shrinking the size of the whole accretion flow. To show the
energy profile for solutions with different $\alpha$, here we
calculate the $\theta$-averaged Bernoulli function $\overline{Be}$.
The Bernoulli function represents the specific total energy in the
accretion flow. A positive value of the Bernoulli function is only a
necessary, not a sufficient, condition for outflow formation
(Abramowicz et al. 2000). However, as our solutions have already
showed outflow structure in the velocity fields, a larger value of
the Bernoulli function indicates that the accretion flow is more
likely to contain stronger outflow. Locally the Bernoulli function
can be written as
\begin{equation}\label{be}
    Be=W+\frac{1}{2}V^2+\Phi=e+\frac{p}{\rho}+\frac{1}{2}V^2+\Phi
    =\frac{GM}{r}\left[\frac{1}{2}(v_r^2(\theta)+v_{\theta}^2(\theta)+
    v_{\phi}^2(\theta))+\frac{\gamma_\mathrm{equ}}{\gamma_\mathrm{equ}-1}\frac{p(\theta)}{\rho{\theta}}-1\right],
\end{equation}
in which $W$ is the specific enthalpy, $V$ is the velocity (all
three components included) and $\Phi$ is the gravitational
potential. So the $\theta$-averaged Bernoulli function can be
written as
\begin{eqnarray}
  \nonumber \overline{Be} &=& \frac{\int_{\theta_\mathrm{b}}^{90\textordmasculine}
    \rho(\theta)\cdot Be \cdot 2\pi r^2 \sin\theta dr\cdot (\pi/180\textordmasculine)d\theta}
    {\int_{\theta_\mathrm{b}}^{90\textordmasculine}
    \rho(\theta)\cdot 2\pi r^2 \sin\theta dr\cdot (\pi/180\textordmasculine)d\theta}
    \\
   &=& \frac{GM}{r}\frac{\int_{\theta_\mathrm{b}}^{90\textordmasculine}
    \rho(\theta)\cdot \left[\frac{1}{2}(v_r^2(\theta)+v_{\theta}^2(\theta)+
    v_{\phi}^2(\theta))+\frac{\gamma_\mathrm{equ}}{\gamma_\mathrm{equ}-1}\frac{p(\theta)}{\rho{\theta}}-1\right] \sin\theta d\theta}
    {\int_{\theta_\mathrm{b}}^{90\textordmasculine} \rho(\theta)\cdot \sin\theta
    d\theta}.
\end{eqnarray}
Figure 18 displays how the $\theta$-averaged Bernoulli function
$\overline{Be}$ changes with variant $\alpha$ for the advection
dominated cases. It can be seen that the positions of the peaks
agree quite well with the bends of line $a_3^\prime$ and line
$a_4^\prime$ in Figure 8.

Figure 19 displays the velocity fields of solutions around the bend
for variant $\alpha$. The values of $\alpha$ are chosen so that the
solutions on both sides of the peaks have similar upper boundaries
($\sim 10\textordmasculine$). It can be seen that the velocities in
the outflow region are more aligned to the $\theta$ direction in the
right panel of each row than those in the left panel. That's because
larger $\alpha$ increases the angular momentum transfer rate, thus
increases the absolute value of $v_r$ near the equatorial plane
where most of the material is accreted. The value of $v_r$ is
negative in the equatorial plane and changes gradually to positive
values in the outflow region under the effect of pressure gradient
and centrifugal force. The absolute value of $v_r$ in the equatorial
plane in the right panel of each row is significantly larger than
that in the left panel, so when pressure gradient doesn't differ
much, the positive values of $v_r$ in the outflow region will be
smaller for larger $\alpha$ and the velocities will be more aligned
to the $\theta$ direction. The solutions corresponding to slim disks
(the second row) show velocities more aligned to the radial
direction in the outflow region than those corresponding to ADAFs
(the first row), due to larger pressure gradient and thus larger
values of $v_r$ in the outflow region.

The bends in solution dependence on $n$, $f$ and
$\gamma_\mathrm{equ}$ have different physical mechanism from that on
$\alpha$. It has been mentioned that, as $n$ increases, $f$
increases or $\gamma_\mathrm{equ}$ decreases, the pressure gradient
increases and drives stronger outflow. However, the outflow rate
cannot exceed the inflow rate. At seen in Figure 17, the ratio of
the outflow rate to the inflow rate for lines $b_4$, $c_2$ and $d_2$
will eventually reach an upper limit at certain values of
corresponding parameters. For solution sets $c_2$ and $d_2$, further
increase in $f$ or further decrease in $\gamma_\mathrm{equ}$ will
continue to increase the radial outflow velocity, while the outflow
rate in the outflow region is almost unchanged, so the size of the
outflow region will shrink. For the solution dependence on $n$, the
bend near $n=1.25$ has similar physical reason as the bends for $f$
and $\gamma_\mathrm{equ}$. However, the size of the accretion flow
has another peak at $n=3/2$, where the flow is always radial (with
rotation) and no outflow is needed. So line $b_4$ in Figure 17
starts to drop as $n$ gets close to 3/2 and line $b_4^\prime$ in
Figure 11 gets to another peak at $n=3/2$.

Figure 20 displays the velocity fields of solutions around the bend
for variant $n$, $f$ and $\gamma_\mathrm{equ}$. It can be seen that,
as $n$ increases (the first row), $f$ increases (the second row) or
$\gamma_\mathrm{equ}$ decreases (the third row), the flow patterns
display similar evolvement for these parameters: at first both the
accretion flow size and the radial outflow velocity increase; after
the corresponding parameter gets past the respective value at the
transition point (the middle panels in all three rows of Figure 20),
the radial outflow velocity continues to increase while the size of
the accretion flow decreases. The velocities in the outflow regions
become more aligned to the radial direction from left to right in
all three rows, due to increased radial outflow velocities.

The labels for solution sets and lines which appear in the text and
figures are summarized in Table 1 along with their corresponding
input parameters.

\begin{center}
\footnotesize {\sc Table 1 \\} {\sc Labels and corresponding
parameters} \vglue 0.2cm
\begin{tabular}{lcccccl}\hline \hline
Label for solution set & Label for lines &$\alpha$ & $n$ & $f$ &
$\gamma_\mathrm{equ}$  \\ \hline
$a_1$   &   $a_1$, $a_1^\prime$   &   0.01 - 1   &   1.3 &   0.01   &   5/3       \\
$a_2$   &   $a_2$, $a_2^\prime$   &   0.01 - 1   &   1.3 &   0.01   &   4/3       \\
$a_3$   &   $a_3$, $a_3^\prime$   &   0.01 - 1   &   1.3 &   1      &   5/3       \\
$a_4$   &   $a_4$, $a_4^\prime$   &   0.01 - 1   &   1.3 &   1      &   4/3       \\
$b_1$   &   $b_1$, $b_1^\prime$   &   0.1   &   0 - 1.5   &   0.01   &   5/3      \\
$b_2$   &   $b_2$, $b_2^\prime$   &   0.1   &   0 - 1.5   &   0.01   &   4/3      \\
$b_3$   &   $b_3$, $b_3^\prime$   &   0.1   &   0 - 1.5   &   1      &   5/3      \\
$b_4$   &   $b_4$, $b_4^\prime$   &   0.1   &   0 - 1.5   &   1      &   4/3      \\
$c_1$   &   $c_1$, $c_1^\prime$   &   0.1   &   1.3   &   0.01 - 1   &   5/3      \\
$c_2$   &   $c_2$, $c_2^\prime$   &   0.1   &   1.3   &   0.01 - 1   &   4/3      \\
$d_1$   &   $d_1$, $d_1^\prime$   &   0.1   &   1.3   &   0.01   &   4/3 - 5/3    \\
$d_2$   &   $d_2$, $d_2^\prime$   &   0.1   &   1.3   &   1      &   4/3 - 5/3    \\
 \hline
\end{tabular}
\end{center}

\section{SUMMARY AND DISCUSSION}

With the self-similar assumptions along the radial direction and
boundary conditions obtained from the reflection symmetry in the
equatorial plane, we are able to solve the Navier-Stokes Equations
along the $\theta$ direction explicitly, and the velocity field is
obtained. The result shows that outflows are common in all kinds of
accretion disk models, which is consistent with some numerical
simulation works (e.g. Ohsuga et al.2005; Ohsuga \& Mineshige 2007;
Ohsuga et al. 2009). Generally the accretion flow consists of three
different regions: an inflow region near the equatorial plane which
contains the largest portion of mass, an outflow region above the
inflow region in which the matter starts escaping the central
accretor in the $r$-direction, and a wind region which contains the
material blowing out from the boundary of the outflow region. The
structure of the inflow and the outflow region can be resolved in
our solutions, while the wind region doesn't obey the self-similar
assumptions and can not be resolved by our calculation. The inflow
region and the wind region are essential to form a steady accretion
flow structure. The outflow region is missing in the cases with
large viscosity parameter $\alpha
> \alpha_\mathrm{c}$. The critical value
$\alpha_\mathrm{c}$ depends on other parameters and is $\sim 0.24$
for ADAFs, $\sim 0.74$ for slim disks and $\sim 0.96$ for gas
pressure dominated SSDs when $n=1.3$.

To compare with the popular analytical models, we calculated
solutions with parameters corresponding to the SSD, the ADAF and the
slim disk models. The solutions corresponding to the SSD model have
relatively thin, quasi-Keplerian disks with very small outflow
regions, and the outflows mostly take the form of wind from the disk
surface. The solutions corresponding to the ADAF and slim disk
models have thick, sub-Keplerian disks. The solution corresponding
to the ADAF model has a moderate outflow region which shrinks
quickly with the increase of $\alpha$, and the outflow mainly takes
the form of wind. However as the upper boundary of the outflow
region is close to the vertical axis, eventually the outflow may
take the form of bipolar jets, though we cannot actually resolve the
explicit structure around the vertical axis. The solutions
corresponding to the slim disk model has a large outflow region in
which most of the outflow takes place, and near the upper boundary
of the outflow region, the material escapes with supersonic
velocity.

Our calculation displays very small outflow regions for input
parameters corresponding to the SSD model, and for these solutions
the inflow region has similar structure with the one-dimensional SSD
model. However the accretion rate inside the disk may not be
conserved along the radius due to mass lost via wind. For the ADAF
and the slim disk cases, our result displays an outflow region too
large to be neglected, and we think it's necessary to consider the
explicit structure in the $\theta$ or $z$ direction to get more
realistic results than the traditional analytical models. It was
proposed in several works (Gu \& Lu, 2007; Jiao et al. 2009) that
the slim disk model has an upper limit of accretion rates, above
which outflows seem to be inevitable. Later in another paper (Jiao
\& Lu 2009), one-dimensional steady transonic global solutions for
slim disks with presumed amount of outflows were calculated, and in
this case the proposed upper limit of accretion rates for slim disks
can be exceeded. Another paper written by Takeuchi et al. (2009)
used the outflow data from two-dimensional radiation-hydrodynamic
(RHD) simulations to construct an one-dimensional steady solution,
and the result shows that the emergent spectra do not sensitively
depend on the amount of mass outflow. However, these models are
still one-dimensional models and they only incorporated the mass
loss effect of outflow into their models, so the results still
remain debatable. We think that two-dimensional solutions
considering the hydrodynamical processes and radiative transfer
along both the radial and the $\theta$ direction are required to
examine the topic, which we are planning to do in our future work.

We also investigated the dependence of the accretion flow structure
on different input parameters respectively. $f$ and
$\gamma_\mathrm{equ}$ (consequently $\beta$) both have some
transition values as discussed in \S{3.2.3} and \S{3.2.4}, at which
the outflow region reaches a maximum size. Generally when $f\gtrsim
0.9$ and $\beta\lesssim 2/3$ (for intrinsic $\gamma=5/3$), the
accretion flow will have a large outflow region and strong outflow,
as the aforementioned solutions corresponding to the slim disk
model. However, in what conditions the transition happens depends on
other input parameters and it won't happen in case of large $\alpha$
due to the existence of the critical value $\alpha_\mathrm{c}$.
Solutions with $n=3/2$ don't contain outflow and they correspond to
the solutions in NY95, which is only applicable to ADAFs. Solutions
with $n<3/2$ have the aforementioned three-region (or two-region for
$\alpha
> \alpha_\mathrm{c}$) structure. Solutions with
$n>3/2$ are not likely to happen in real cases.

Finally, we would like to mention some caveats in our work. The
solutions obtained in this paper are steady solutions, and thus it
is not guaranteed that they are stable solutions. These solutions do
not cover the whole space, so for the unresolved region near the
axis, we know nothing about the detailed structure other than that
it contains the material blowing out of the upper boundary in the
form of wind. Our work is based on the simple $\alpha p$
prescription of viscosity, which is not bad according to recent MHD
simulation works(Hirose et al. 2009; Ohsuga et al. 2009), especially
when the flow is steady. However, it may be necessary to consider
other components of the viscous stress tensor for solutions with
large $v_r$. We also used the advective factor $f$ to simplify the
energy equation, and assumed $f$ to be a constant. Realistically $f$
should vary with both $\theta$ and $r$, which could be improved by
considering the details of radiative transfer. We used Newtonian
gravitational potential in our work, and it needs to be corrected
when applied to study the structure close to the innermost stable
circular orbit. In our calculation we didn't consider the effects of
convection (e.g. Stone et al. 1999; Igumenshchev \& Abramowicz 2000;
Yuan \& Bu 2010) or magnetic field (e.g. Blandford \& Payne 1982;
Naso \& Miller 2010), which may be important in studying the
accretion flow and the generation of outflows. Finally, it is
necessary to abandon the self-similar assumptions if we want to
investigate the disk and outflow structure more quantitatively and
precisely. These caveats will be improved in our future work.

\acknowledgments

We thank Fu-Guo Xie for discussions and suggestions. We are also
grateful to Marek Abramowicz, Feng Yuan, Wei-Min Gu and Li Xue for
helpful comments. We also thank the referee for his/her constructive
suggestions. This work was supported by the NSFC grant (No.
11033001) and the 973 program (No. 2007CB815405).

\clearpage
\begin{figure}
\plotone{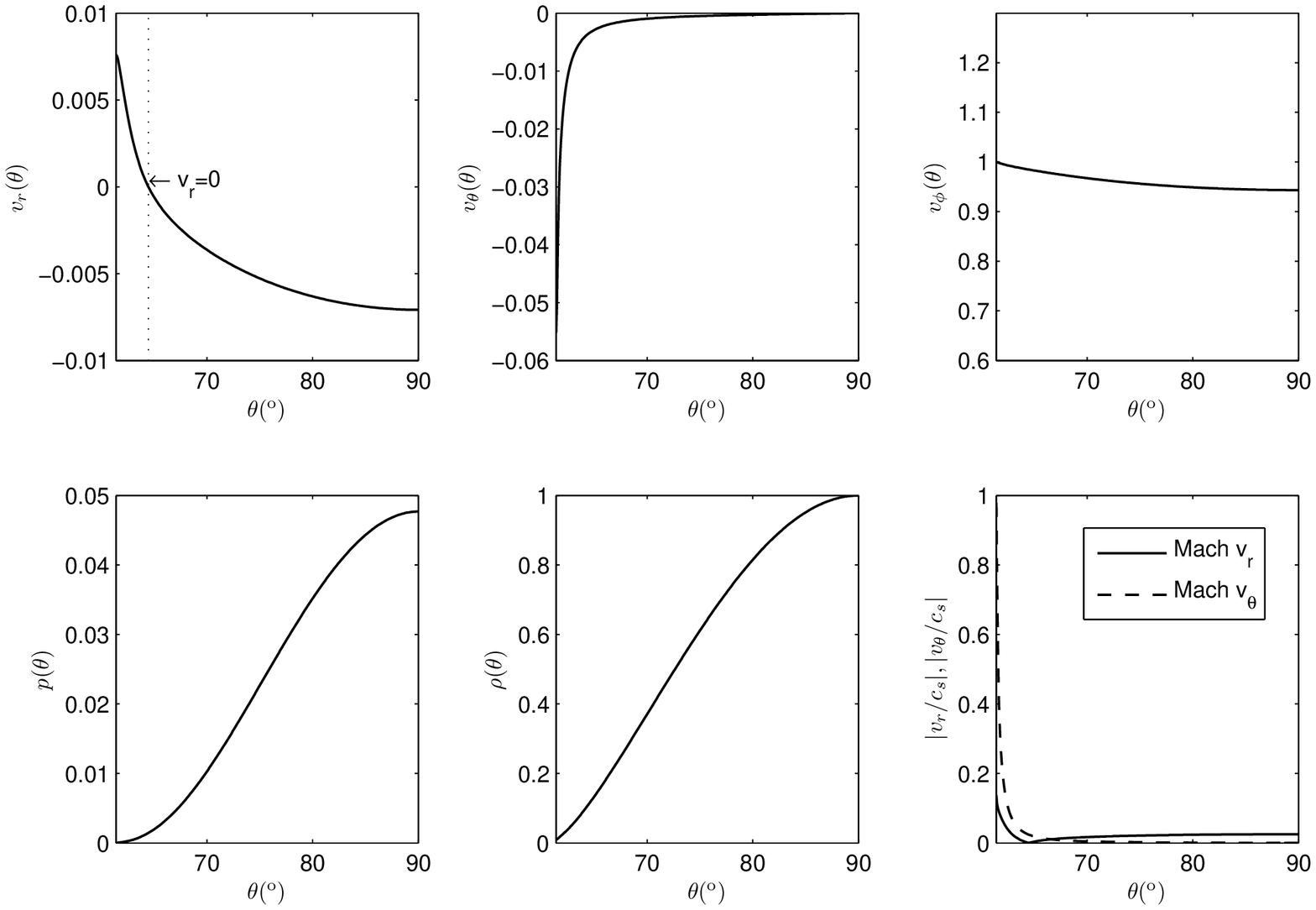} \caption{The solution corresponding to the gas
pressure dominated region of the SSD model. Here $\alpha=0.1$,
$n=1.3$, $f=0.01$ and $\gamma_\mathrm{equ}=5/3$, which correspond to
gas pressure dominated monatomic ideal gas.} \label{fig1}
\end{figure}

\begin{figure}
\plotone{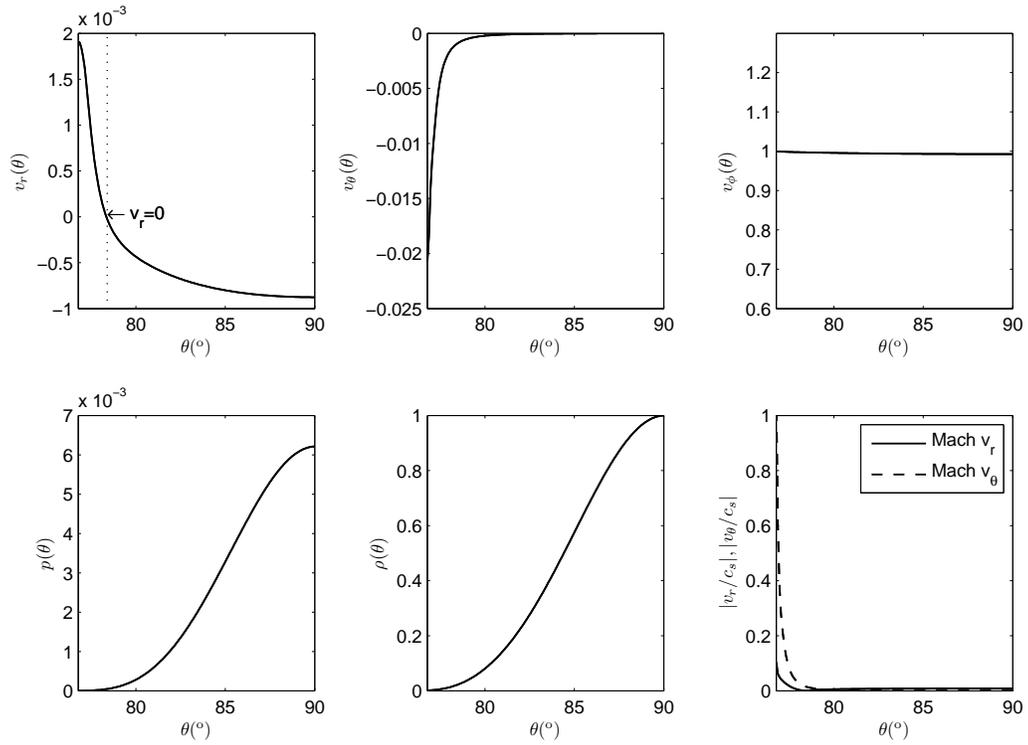} \caption{The solution corresponding to the
radiation pressure dominated region of the SSD model. Here
$\alpha=0.1$, $n=1.3$, $f=0.01$ and $\gamma_\mathrm{equ}=4/3$, which
correspond to radiation pressure dominated monatomic ideal gas.}
\label{fig2}
\end{figure}

\begin{figure}
\plotone{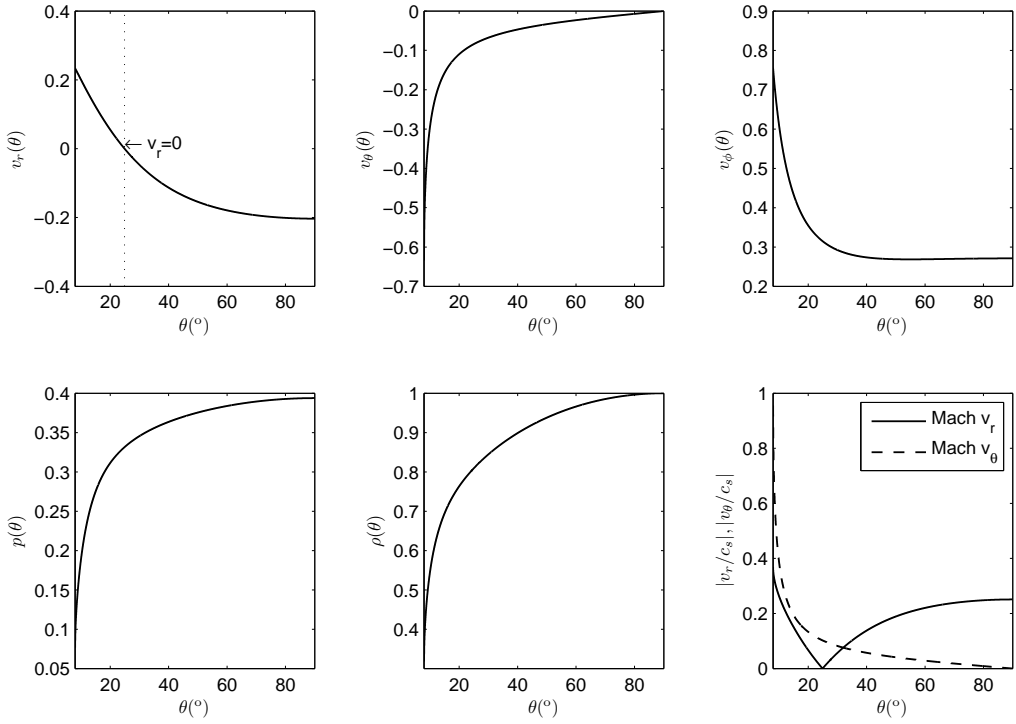} \caption{The solution corresponding to the ADAF
model. Here $\alpha=0.1$, $n=1.3$, $f=1$ and
$\gamma_\mathrm{equ}=5/3$, which correspond to gas pressure
dominated monatomic ideal gas.} \label{fig3}
\end{figure}

\begin{figure}
\plotone{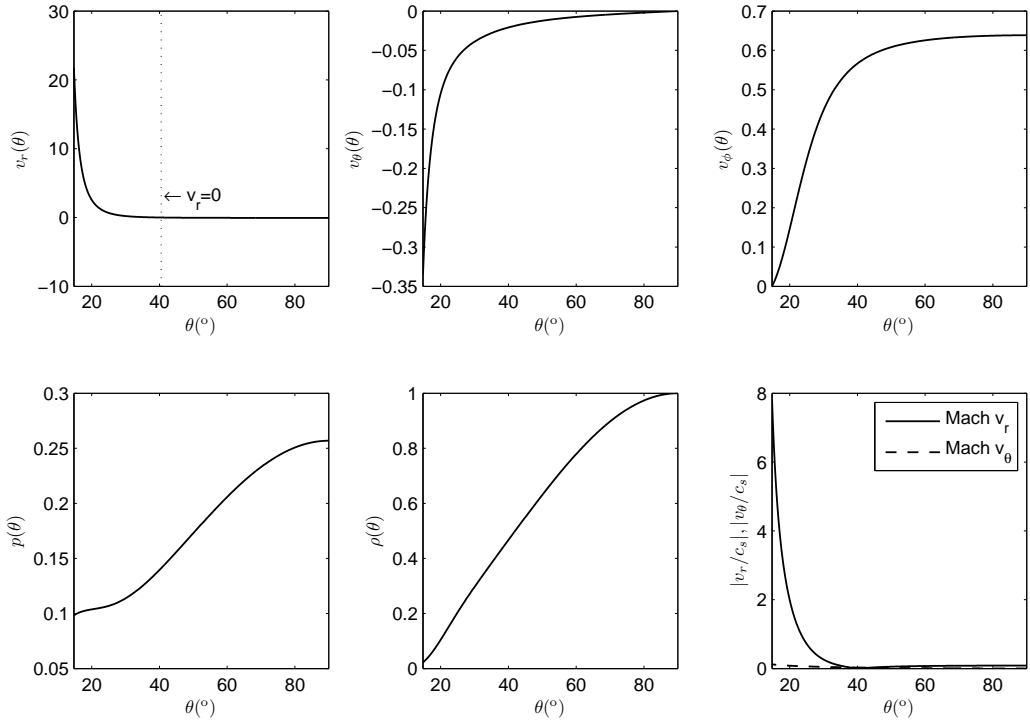} \caption{The solution corresponding to the slim
disk model. Here $\alpha=0.1$, $n=1.3$, $f=1$ and
$\gamma_\mathrm{equ}=4/3$, which correspond to radiation pressure
dominated monatomic ideal gas.}\label{fig4}
\end{figure}

\begin{figure}
\plotone{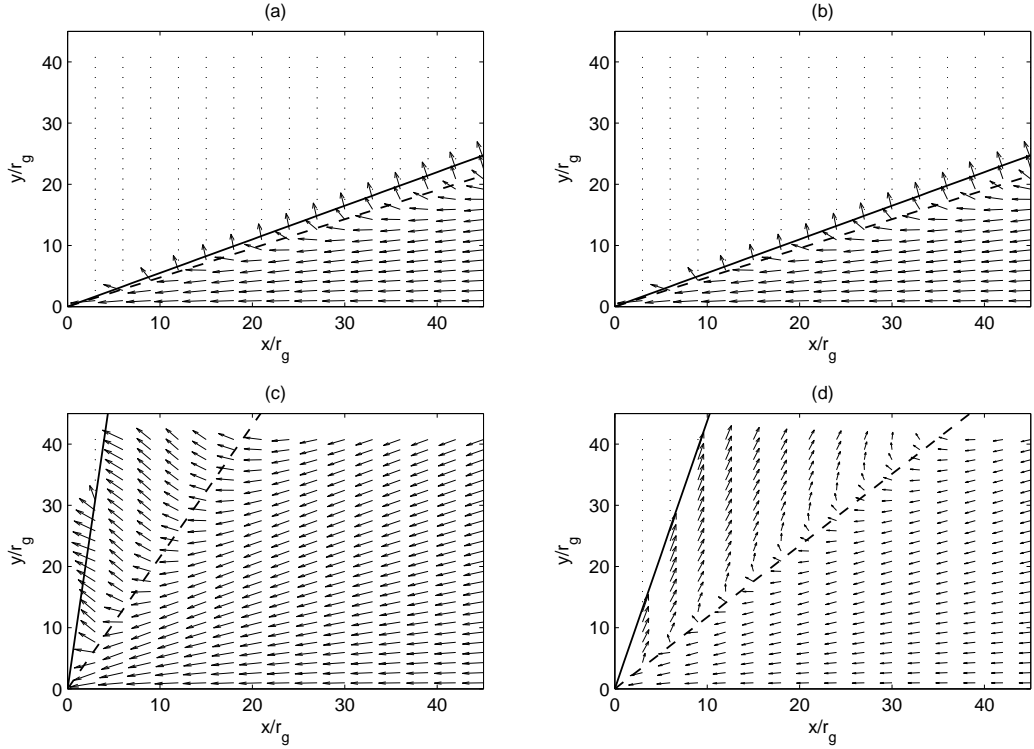} \caption{The velocity fields of the four typical
solutions. Figures (a), (b), (c) and (d) correspond to input
parameters of Figures 1-4 respectively. The lengths of arrows
indicate the absolute values of the vector
$\vec{v_r}(\theta)$+$\vec{v_\theta}(\theta)$, which are scaled
logarithmically. The solid lines correspond to the inclination
$\theta_\mathrm{b}$, while the dashed lines correspond to the
inclination $\theta_0$.} \label{fig5}
\end{figure}

\begin{figure}
\plotone{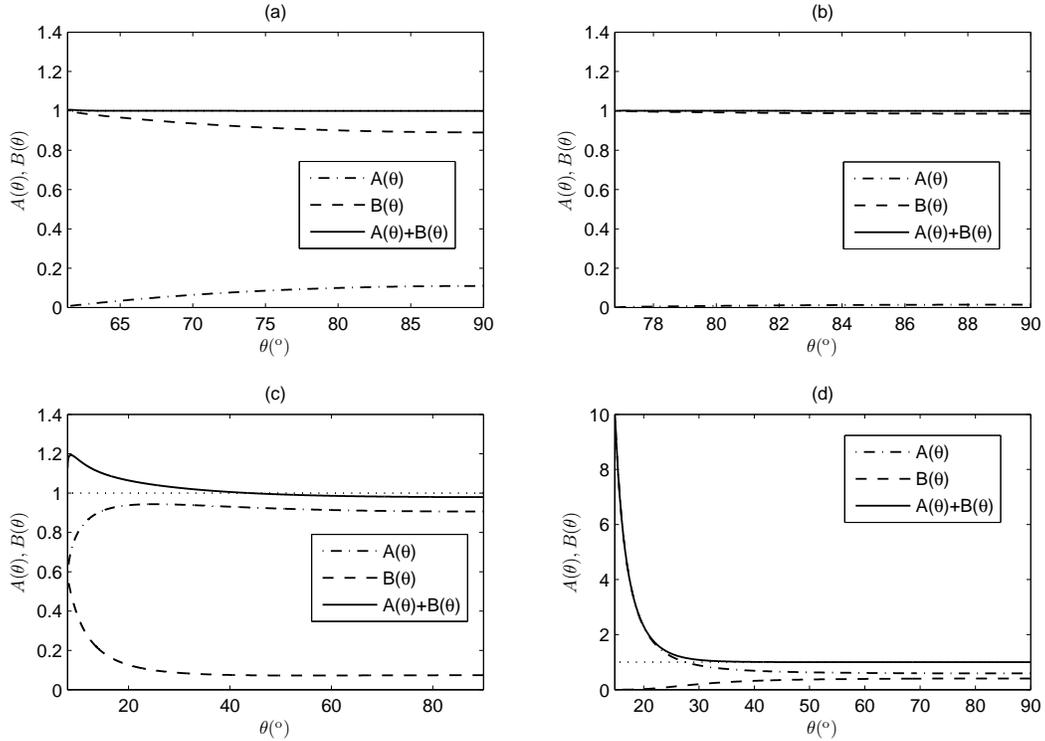} \caption{The radial components of the pressure
gradient, the centrifugal force and the gravitational force for the
four typical solutions. Figures (a), (b), (c) and (d) correspond to
input parameters of Figures 1-4 respectively. The dotted lines
correspond to the gravitational force which is scaled as 1; the
dashed lines correspond to the radial component of the centrifugal
force; the dot-dashed lines correspond to the radial component of
the pressure gradient; and the solid lines correspond to the sum of
the radial components of the centrifugal force and the pressure
gradient, which drives the outflow.}\label{fig6}
\end{figure}

\begin{figure}
\plotone{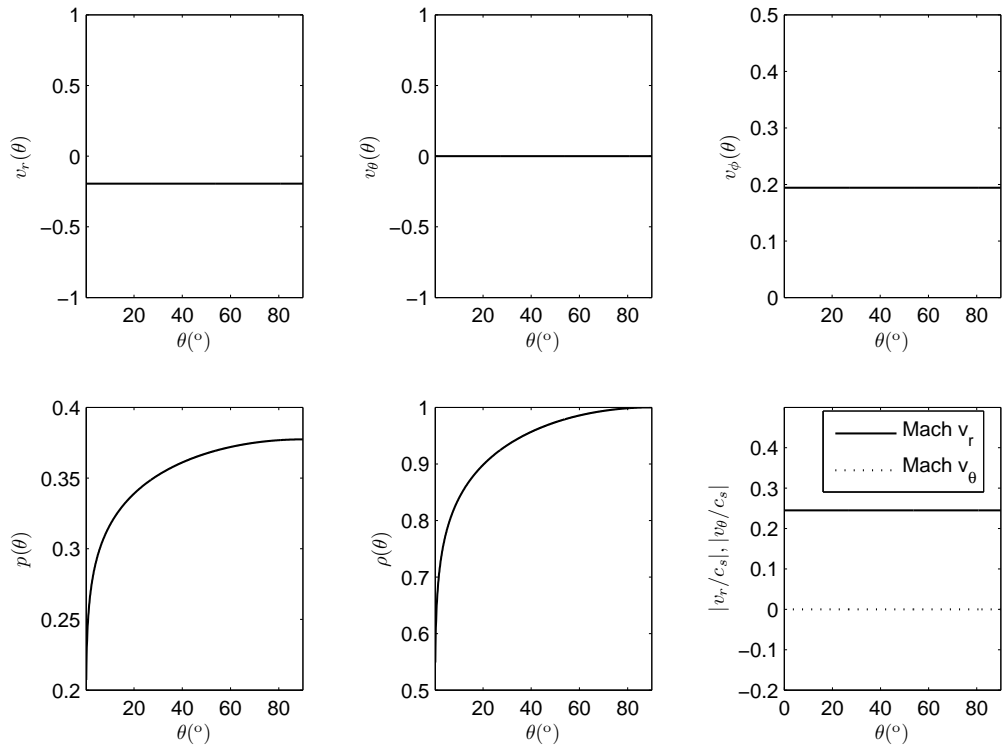} \caption{The solution corresponding to
$\epsilon\prime=0.1$ in NY95. Here $\alpha=0.1$, $n=1.5$, $f=1$ and
$\gamma_\mathrm{equ}=1.6061$.}\label{fig7}
\end{figure}

\begin{figure}
\plotone{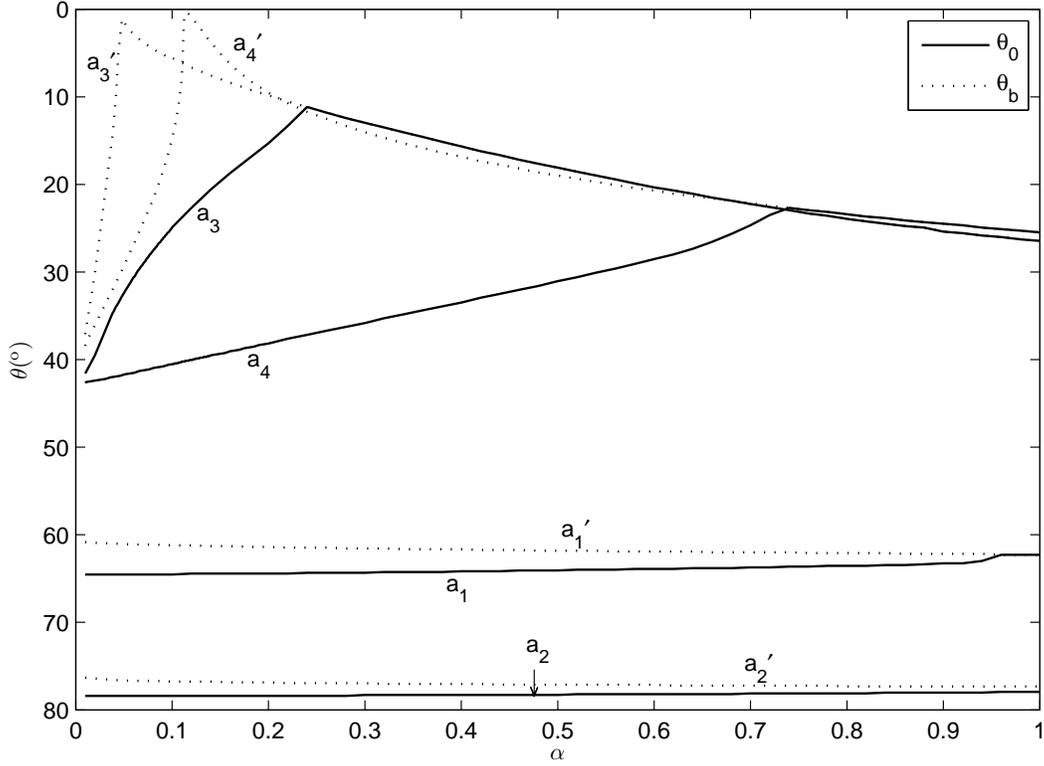} \caption{Solution dependence on $\alpha$. Solid
lines $a_1$, $a_2$, $a_3$ and $a_4$ represent the upper boundaries
of the inflow region, and dotted lines $a_1^\prime$, $a_2^\prime$,
$a_3^\prime$ and $a_4^\prime$ represent the boundaries of the whole
inflow/outflow region. All the lines correspond to $n=1.3$. Lines
$a_1$ and $a_1^\prime$ correspond to $f=0.01$ and
$\gamma_\mathrm{equ}=5/3$; lines $a_2$ and $a_2^\prime$ correspond
to $f=0.01$ and $\gamma_\mathrm{equ}=4/3$; lines $a_3$ and
$a_3^\prime$ correspond to $f=1$ and $\gamma_\mathrm{equ}=5/3$;
lines $a_4$ and $a_4^\prime$ correspond to $f=1$ and
$\gamma_\mathrm{equ}=4/3$.}\label{ca}
\end{figure}

\begin{figure}
\plotone{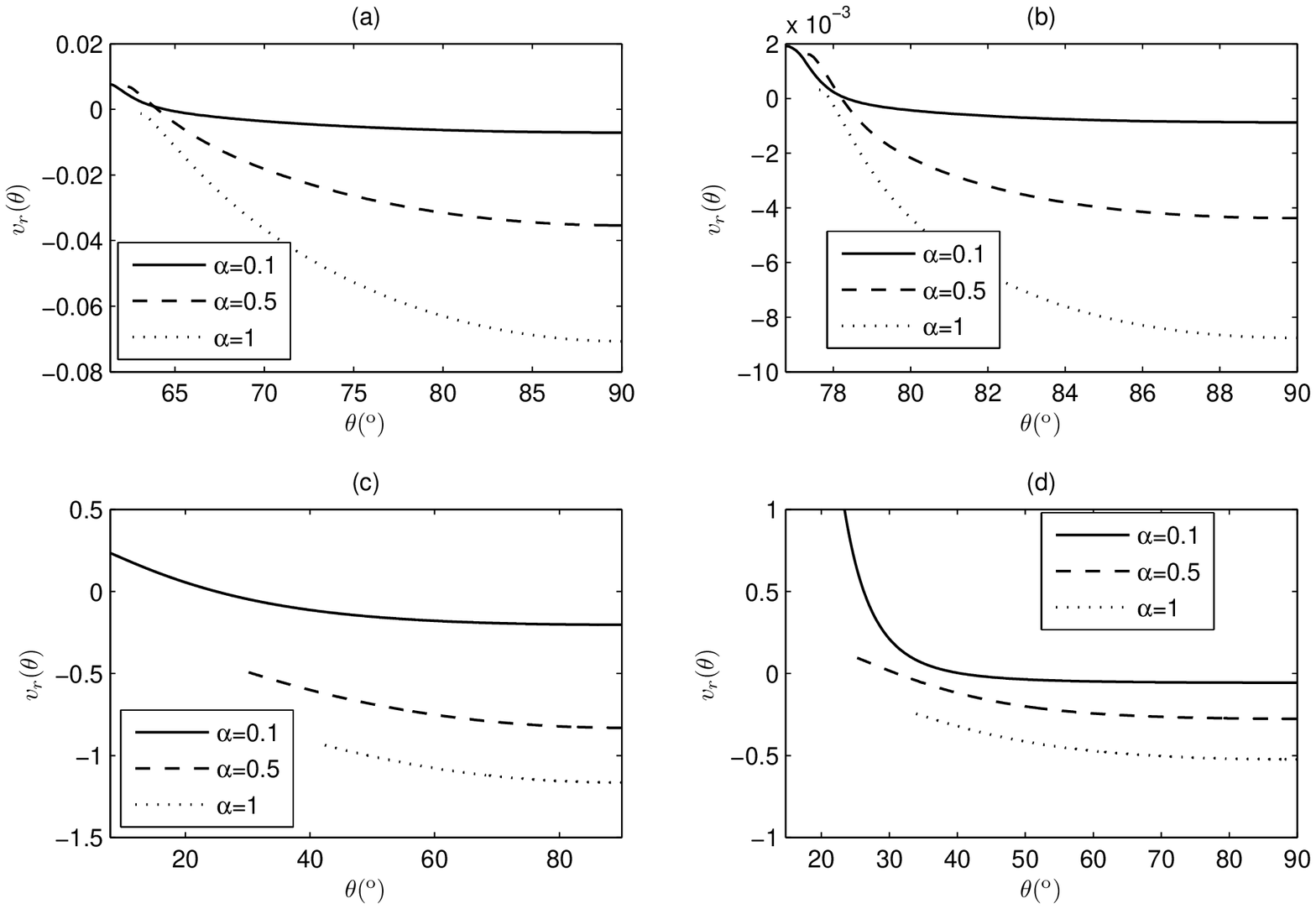} \caption{The distribution of $v_r(\theta)$ along
the $\theta$-direction for solutions with different $\alpha$. Figure
(a) corresponds to $n=1.3$, $f=0.01$ and $\gamma_\mathrm{equ}=5/3$;
Figure (b) corresponds to $n=1.3$, $f=0.01$ and
$\gamma_\mathrm{equ}=4/3$; Figure (c) corresponds to $n=1.3$, $f=1$
and $\gamma_\mathrm{equ}=5/3$; Figure (d) corresponds to $n=1.3$,
$f=1$ and $\gamma_\mathrm{equ}=4/3$. The solid, dashed and dotted
lines correspond to $\alpha=$ 0.1, 0.5 and 1, respectively.}
\label{vr1}
\end{figure}

\begin{figure}
\plotone{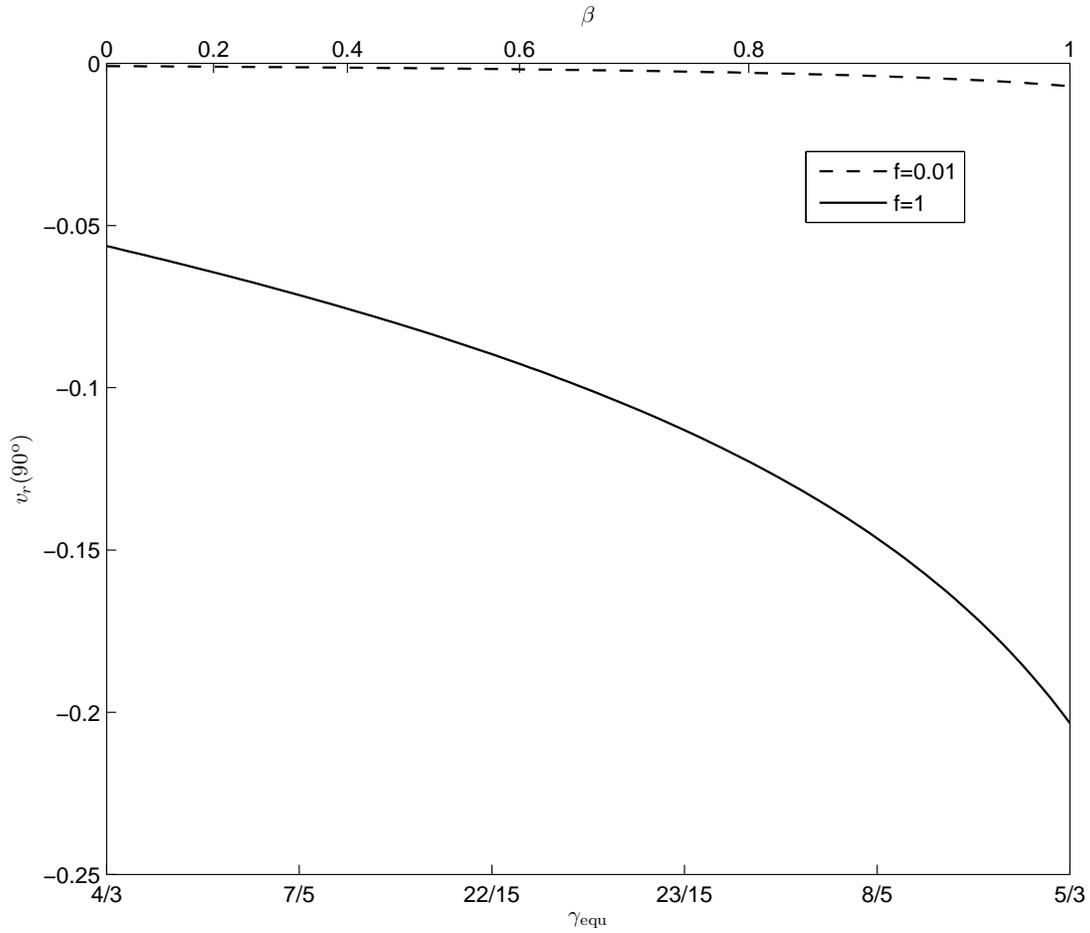} \caption{The equatorial values of $v_r(\theta)$
corresponding to different $\gamma_\mathrm{equ}$. All the solutions
correspond to $\alpha=0.1$ and $n=1.3$. The solid line corresponds
to $f=1$, while the dashed line corresponds to $f=0.01$. $\beta$ on
the top axis corresponds to $\gamma_\mathrm{equ}$ on the bottom axis
and is calculated based on $\gamma=5/3$.} \label{vr_equ}
\end{figure}

\begin{figure}
\plotone{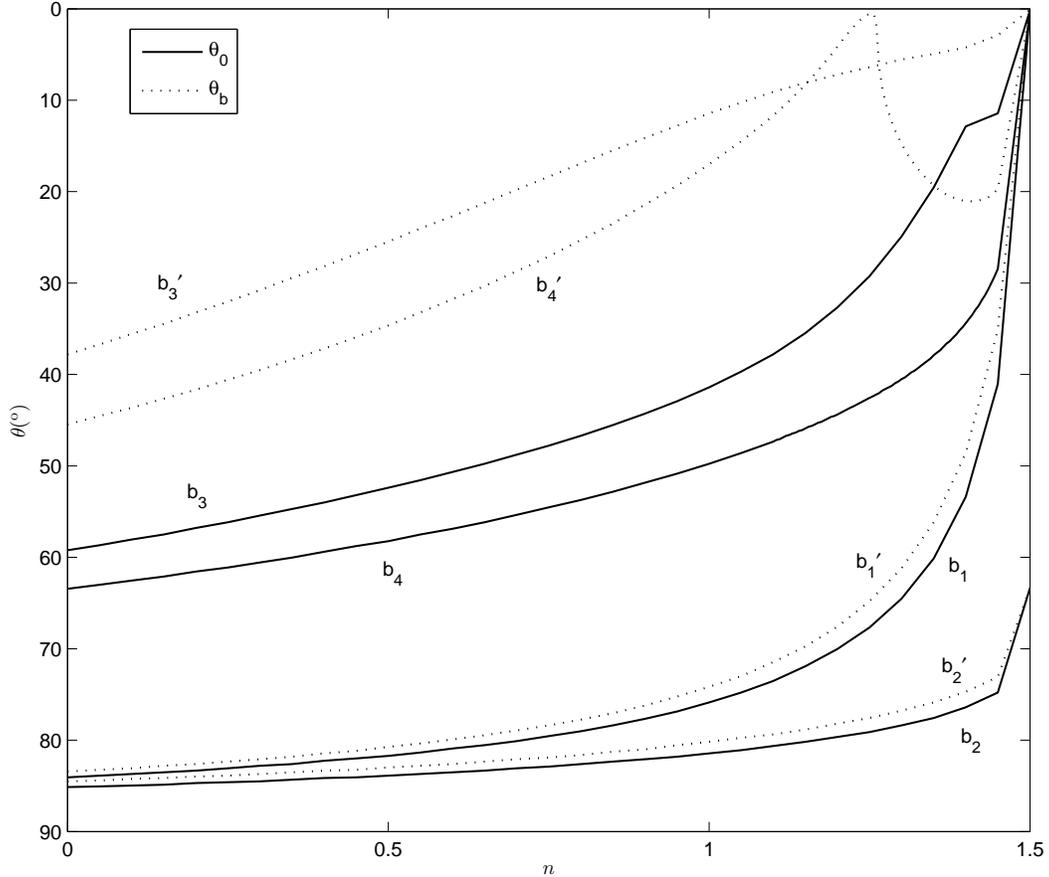} \caption{Solution dependence on $n$. Solid lines
$b_1$, $b_2$, $b_3$ and $b_4$ represent the upper boundary of the
inflow region, and dotted lines $b_1^\prime$, $b_2^\prime$,
$b_3^\prime$ and $b_4^\prime$ represents the boundaries of the whole
inflow/outflow region. All the lines correspond to $\alpha=0.1$.
Lines $b_1$ and $b_1^\prime$ correspond to $f=0.01$ and
$\gamma_\mathrm{equ}=5/3$; lines $b_2$ and $b_2^\prime$ correspond
to $f=0.01$ and $\gamma_\mathrm{equ}=4/3$; lines $b_3$ and
$b_3^\prime$ correspond to $f=1$ and $\gamma_\mathrm{equ}=5/3$;
lines $b_4$ and $b_4^\prime$ correspond to $f=1$ and
$\gamma_\mathrm{equ}=4/3$.}\label{cn}
\end{figure}

\begin{figure}
\plotone{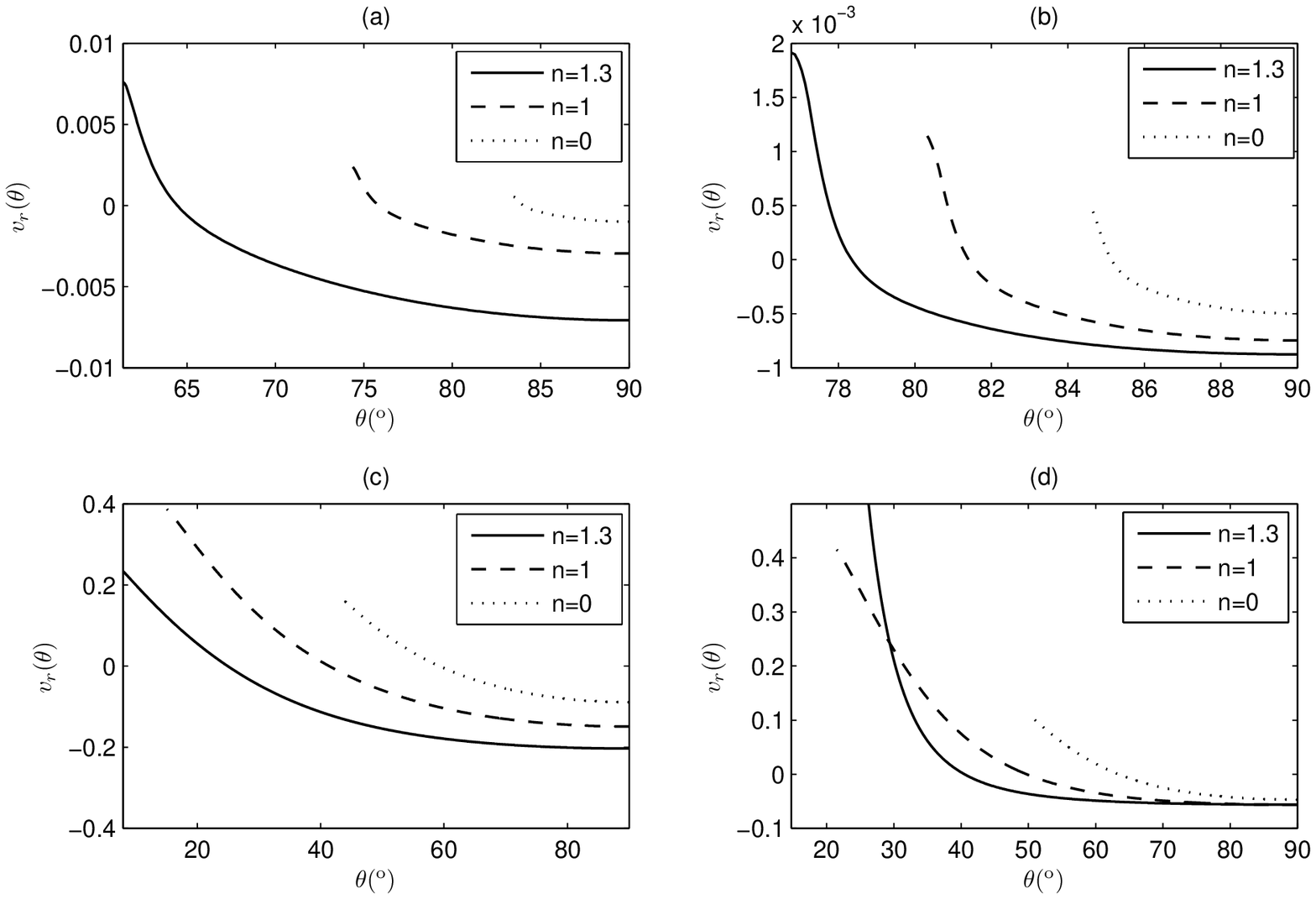} \caption{The distribution of $v_r(\theta)$ along
the $\theta$-direction for solutions with different $n$. Figure (a)
corresponds to $\alpha=0.1$, $f=0.01$ and $\gamma_\mathrm{equ}=5/3$;
Figure (b) corresponds to $\alpha=0.1$, $f=0.01$ and
$\gamma_\mathrm{equ}=4/3$; Figure (c) corresponds to $\alpha=0.1$,
$f=1$ and $\gamma_\mathrm{equ}=5/3$; Figure (d) corresponds to
$\alpha=0.1$, $f=1$ and $\gamma_\mathrm{equ}=4/3$. The solid, dashed
and dotted lines correspond to $n=$ 1.3, 1 and 0, respectively.}
\label{vr2}
\end{figure}

\begin{figure}
\plotone{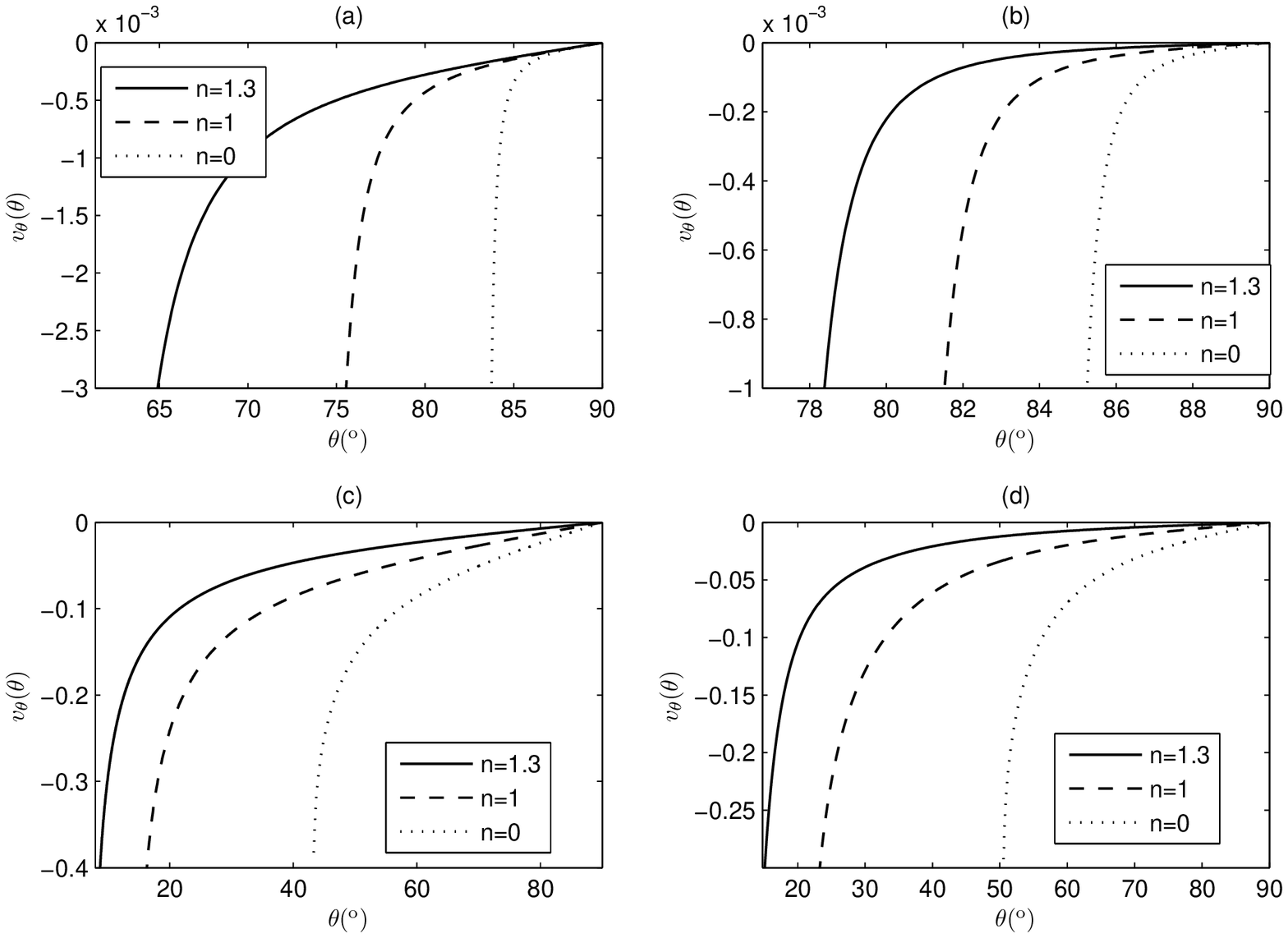} \caption{The distribution of $v_\theta(\theta)$
along the $\theta$-direction for solutions with different $n$.
Figure (a) corresponds to $\alpha=0.1$, $f=0.01$ and
$\gamma_\mathrm{equ}=5/3$; Figure (b) corresponds to $\alpha=0.1$,
$f=0.01$ and $\gamma_\mathrm{equ}=4/3$; Figure (c) corresponds to
$\alpha=0.1$, $f=1$ and $\gamma_\mathrm{equ}=5/3$; Figure (d)
corresponds to $\alpha=0.1$, $f=1$ and $\gamma_\mathrm{equ}=4/3$.
The solid, dashed and dotted lines correspond to $n=$ 1.3, 1 and 0,
respectively.} \label{vt2}
\end{figure}

\begin{figure}
\plotone{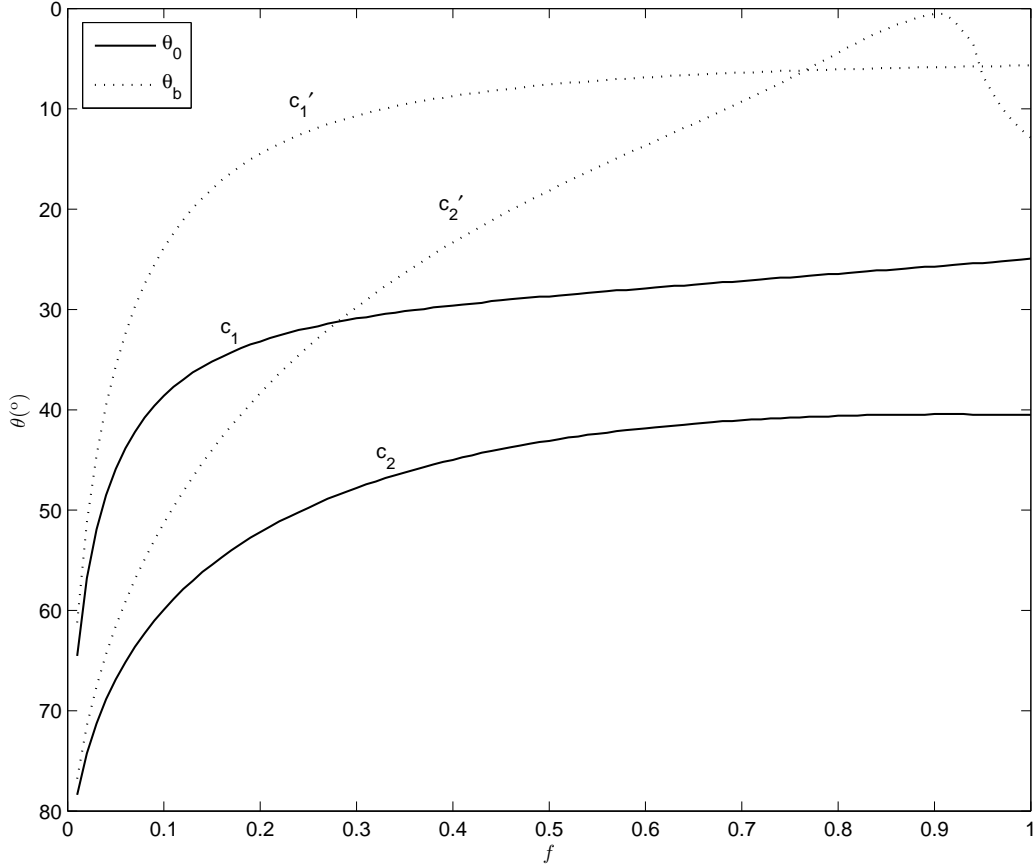} \caption{Solution dependence on $f$. Solid lines
$c_1$ \& $c_2$ represent the upper boundary of the inflow region,
and dotted lines $c_1^\prime$ \& $c_2^\prime$ represents the outer
boundary of the whole accretion flow. All the lines correspond to
$\alpha=0.1$ \& $n=1.3$. Lines $c_1$ and $c_1^\prime$ correspond to
$\gamma_\mathrm{equ}=5/3$; lines $c_2$ and $c_2^\prime$ correspond
to $\gamma_\mathrm{equ}=4/3$.}\label{cf}
\end{figure}

\begin{figure}
\plotone{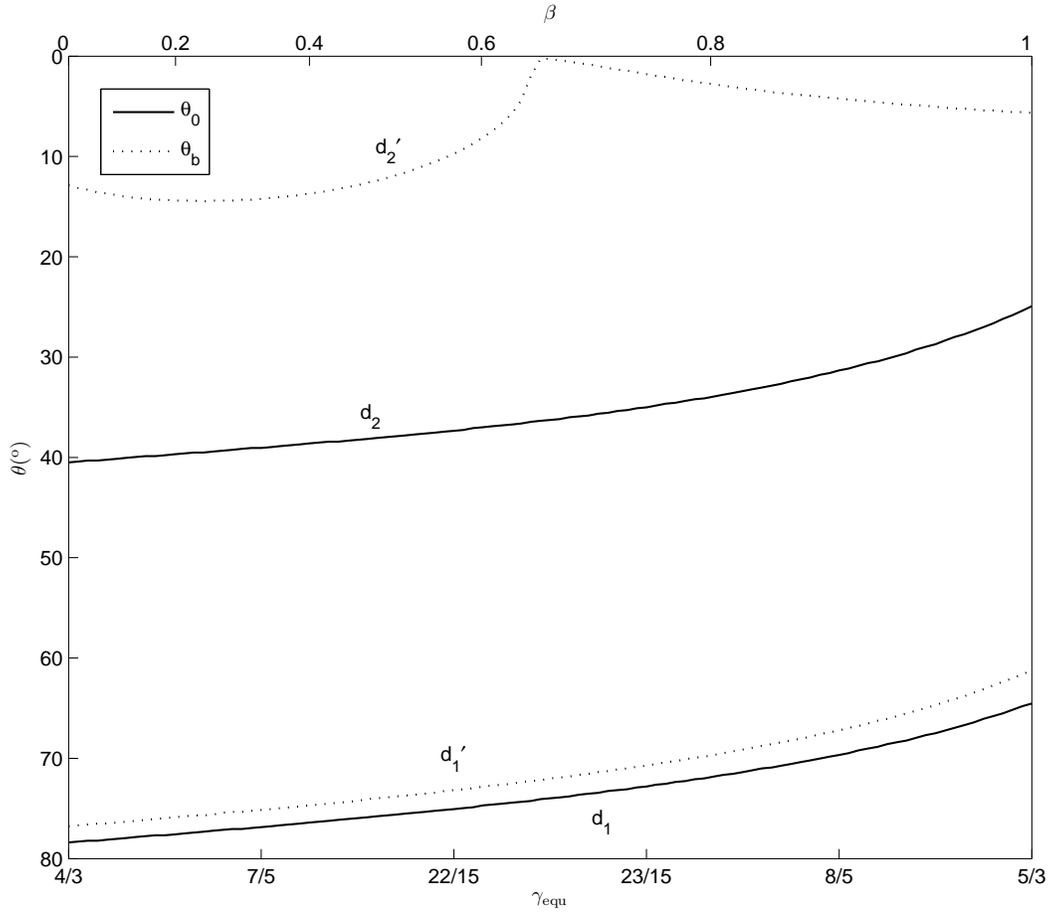} \caption{Solution dependence on
$\gamma_\mathrm{equ}$. Solid lines $d_1$ \& $d_2$ represent the
upper boundary of the inflow region, and dotted lines $d_1^\prime$
\& $d_2^\prime$ represent the upper boundary of the whole
inflow/outflow region. All the lines correspond to $\alpha=0.1$ \&
$n=1.3$. Lines $d_1$ and $d_1^\prime$ correspond to $f=0.01$; lines
$d_2$ and $d_2^\prime$ correspond to $f=1$. $\beta$ on the top axis
corresponds to $\gamma_\mathrm{equ}$ on the bottom axis and is
calculated based on $\gamma=5/3$. }\label{cg}
\end{figure}

\begin{figure}
\plotone{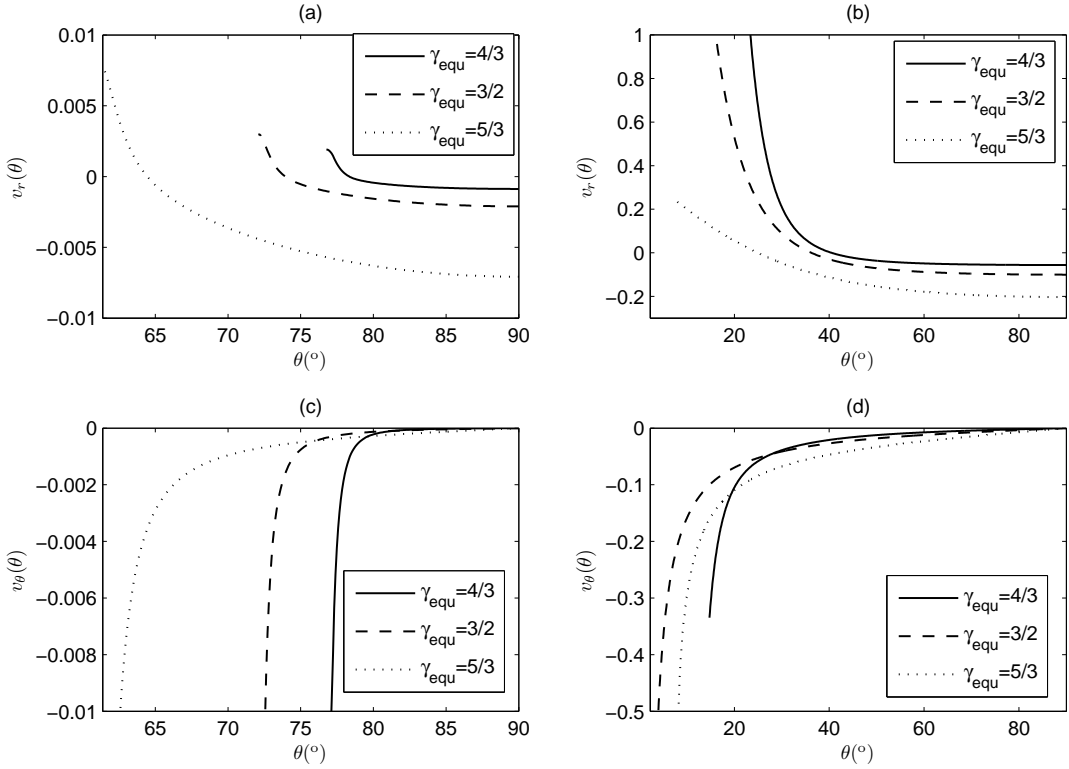} \caption{The $v_r$ and $v_\theta$ distributions
along the $\theta$-direction for solutions with different
$\gamma_\mathrm{equ}$. Figures (a) \& (c) correspond to solutions
with $n=1.3$, $\alpha=0.1$ and $f=0.01$, while Figures (b) \& (d)
correspond to solutions with $n=1.3$, $\alpha=0.1$ and $f=1$. The
solid, dashed and dotted lines correspond to $\gamma_\mathrm{equ}=$
4/3, 3/2 and 5/3, respectively.} \label{vrt4}
\end{figure}

\begin{figure}
\plotone{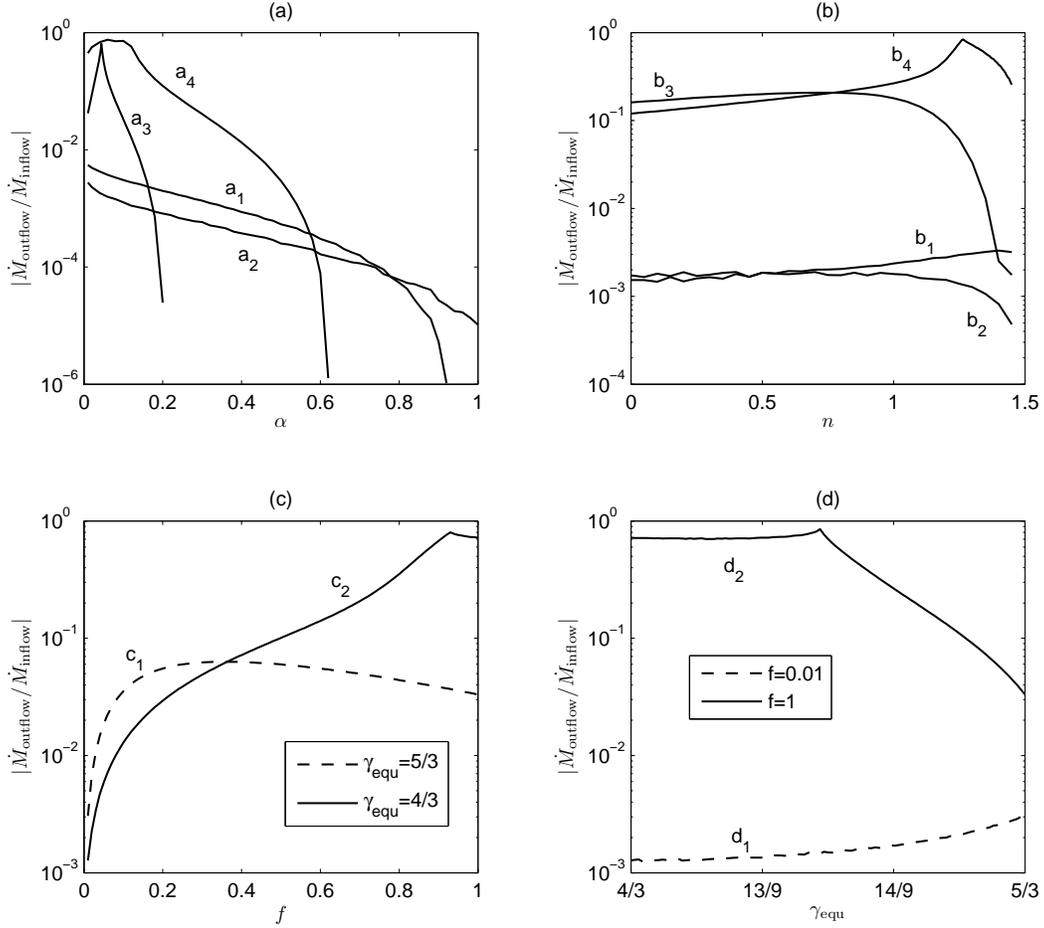} \caption{The ratio of the outflow rate to the
inflow rate. The vertical axes show the absolute values of the
ratio, and the horizontal axes correspond to different input
parameters. The lines in the top left panel correspond to the lines
in Figure 8. The lines in the top right panel correspond to the
lines in Figure 11. The lines in the bottom left panel correspond to
the lines in Figure 14. The lines in the bottom right panel
correspond to the lines in Figure 15. Certain parts of some lines
are missing due to that the solutions there don't contain an outflow
region.} \label{ratio1}
\end{figure}

\begin{figure}
\plotone{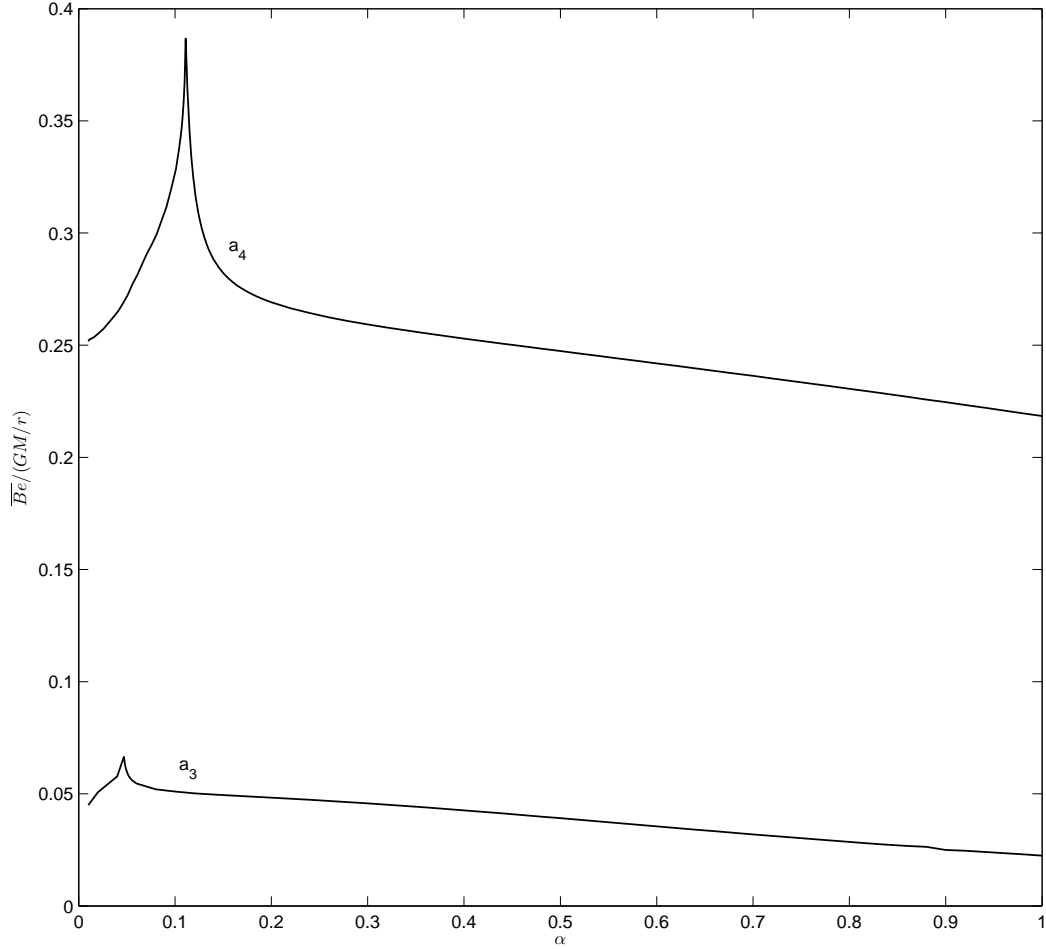} \caption{The $\theta$-averaged Bernoulli function
for variant $\alpha$. Line $a_3$ corresponds to solutions with
$f=1$, $\gamma_\mathrm{equ}=5/3$ and $n=1.3$. Line $a_4$ corresponds
to solutions with $f=1$, $\gamma_\mathrm{equ}=4/3$ and $n=1.3$.}
\label{be}
\end{figure}

\clearpage

\begin{figure}
\plotone{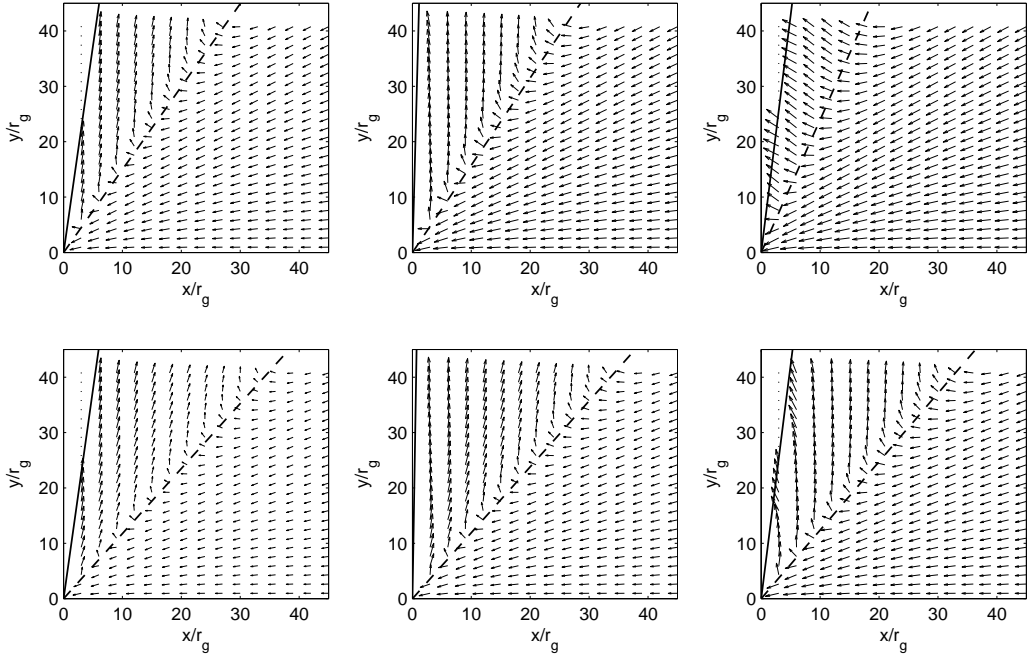} \caption{The velocity field patterns around the
bends for variant $\alpha$. The first row of figures corresponds to
the bend of line $a_3^\prime$ in Figure 8, with $\alpha=$ 0.043,
0.05 and 0.12 from left to right respectively, and other parameters
$f=1$, $\gamma_\mathrm{equ}=5/3$ and $n=1.3$. The second row of
figures corresponds to the bend of line $a_4^\prime$ in Figure 8,
with $\alpha=$ 0.106, 0.12 and 0.165 from left to right
respectively, and other parameters $f=1$, $\gamma_\mathrm{equ}=4/3$
and $n=1.3$. The lengths of arrows indicate the absolute values of
the vector $\vec{v_r}(\theta)$+$\vec{v_\theta}(\theta)$, which are
scaled logarithmically. The solid lines correspond to the
inclination $\theta_\mathrm{b}$, while the dashed lines correspond
to the inclination $\theta_0$.} \label{flow patterns 1}
\end{figure}

\begin{figure}
\plotone{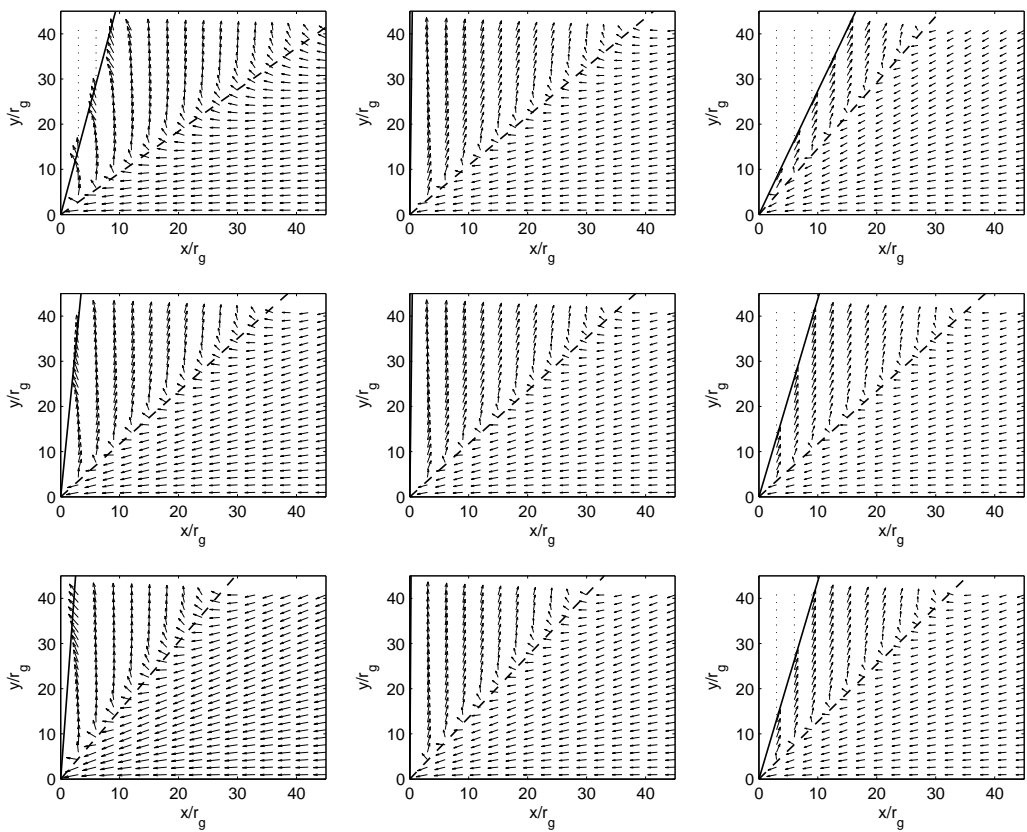} \caption {The velocity field patterns around the
bends for variant $n$, $f$ and $\gamma_\mathrm{equ}$. The first row
of figures corresponds to the bend of line $b_4^\prime$ in Figure
11, with $n=$ 1.1, 1.25 and 1.4 from left to right respectively, and
other parameters $\alpha=0.1$, $f=1$ and $\gamma_\mathrm{equ}=4/3$.
The second row of figures corresponds to the bend of line
$c_2^\prime$ in Figure 14, with $f=$ 0.8, 0.9 and 1 from left to
right respectively, and other parameters $\alpha=0.1$,
$\gamma_\mathrm{equ}=4/3$ and $n=1.3$. The third row of figures
corresponds to the bend of line $d_2^\prime$ in Figure 15, with
$\gamma_\mathrm{equ}=$ 47/30, 3/2 and 43/30 from left to right
respectively, and other parameters $\alpha=0.1$, $f=1$ and $n=1.3$.
The lengths of arrows indicate the absolute values of the vector
$\vec{v_r}(\theta)$+$\vec{v_\theta}(\theta)$, which are scaled
logarithmically. The solid lines correspond to the inclination
$\theta_\mathrm{b}$, while the dashed lines correspond to the
inclination $\theta_0$.} \label{flow patterns 2}
\end{figure}

\end{document}